\documentclass[12pt]{article}

  \usepackage{graphicx}
  \usepackage{epsfig}
  \usepackage{amssymb}
  \usepackage{subfigure}
  \usepackage{amsmath} 

\evensidemargin - .5truecm
\allowdisplaybreaks[1]

\def \lsim
{\raisebox{-3pt}{$\>\stackrel{<}{\scriptstyle\sim}\>$}}
\def \gsim
{\raisebox{-3pt}{$\>\stackrel{>}{\scriptstyle\sim}\>$}}

\begin{document}

\newcommand{\CC}{{\mathbb C}}
\newcommand{\RR}{{\mathbb R}}
\newcommand{\ZZ}{{\mathbb Z}}
\newcommand{\QQ}{{\mathbb Q}}
\newcommand{\NN}{{\mathbb N}}
\newcommand{\beq}{\begin{equation}}
\newcommand{\eeq}{\end{equation}}
\newcommand{\beal}{\begin{align}}
\newcommand{\eeal}{\end{align}}
\newcommand{\nn}{\nonumber}
\newcommand{\bea}{\begin{eqnarray}}
\newcommand{\eea}{\end{eqnarray}}
\newcommand{\ba}{\begin{array}}
\newcommand{\ea}{\end{array}}
\newcommand{\bfig}{\begin{figure}}
\newcommand{\efig}{\end{figure}}
\newcommand{\bc}{\begin{center}}
\newcommand{\ec}{\end{center}}

\newenvironment{appendletterA}
{
  \typeout{ Starting Appendix \thesection }
  \setcounter{section}{0}
  \setcounter{equation}{0}
  \renewcommand{\theequation}{A\arabic{equation}}
 }{
  \typeout{Appendix done}
 }
\newenvironment{appendletterB}
 {
  \typeout{ Starting Appendix \thesection }
  \setcounter{equation}{0}
  \renewcommand{\theequation}{B\arabic{equation}}
 }{
  \typeout{Appendix done}
 }

%
%
%
%

\begin{titlepage}
\nopagebreak

\renewcommand{\thefootnote}{\fnsymbol{footnote}}
\vskip 2cm
\begin{center}
\boldmath

{\Large\bf Winter Model at Finite Volume}

\unboldmath
\vskip 1.cm
{\large  U.~G.~Aglietti}
\vskip .2cm
{\it Dipartimento di Fisica, Universit\`a di Roma ``La Sapienza''}
\end{center}
\vskip 0.5cm

\begin{abstract}
We study Winter or $\delta$-shell model at finite volume (length),
describing a small resonating cavity
weakly-coupled to a large one.
For generic values of the coupling,
a resonance of the usual model corresponds, 
in the finite-volume case,
to a compression of the spectral lines;
for specific values of the coupling,
a resonance corresponds instead to a
degenerate or a quasi-degenerate doublet.
A secular term of the form $g^3 N$
occurs in the perturbative expansion
of the momenta (or of the energies)
of the particle at third order in $g$, 
where $g$ is the coupling 
among the cavities and $N$ is the
ratio of the length of the large cavity
over the length of the small one.
These secular terms, which tend
to spoil the convergence of the
perturbative series in the large
volume case, $N \gg 1$, are resummed 
to all orders in $g$ by means of standard
multi-scale methods.
The resulting improved perturbative expansions
provide a rather complete analytic description
of resonance dynamics at finite volume.

\vskip .4cm

\end{abstract}
\vfill
\end{titlepage}    

\setcounter{footnote}{0}

\newpage

\tableofcontents

\newpage


\section{Introduction}

According to the superposition principle 
of quantum mechanics, an arbitrary
linear combination of wavefunctions
of a physical system describes a
possible, i.e. admissible, state of the latter.
Whether such a state is physically realizable or not,
it just depends on experimental facilities
--- ultimately on technological ability.
In general, the superposition principle
implies a radical departure from classical
mechanics, where particles are completely
localized in space at any time \cite{Dirac}.
Stringent checks on the linear structure of quantum mechanics
are provided by the flavor oscillation phenomena
in neutral kaon and beauty mesons \cite{tdlee}; 
a small non linearity in the time-evolution
of these systems would produce higher-order harmonics 
in the wavefunction (typically second and third-order ones), 
which have never been observed.
Historically, quantum mechanics was created to describe the stationary states 
--- as well as the transitions --- of atoms and molecules 
and it  was thought as the correct general theory of micro-systems, 
in contrast  to classical mechanics, relegated to describe macroscopic systems only.
However, as well known, there is not any spatial or temporal upper limit to the coherence phenomena implied by the superposition
principle.
In other words, while in quantum mechanics 
--- as formulated nowadays --- 
there exists, in comparison to classical physics, 
a new scale controlling the phenomena at small distances
--- namely the Planck constant or "quantum of action" $h$ ---, 
there is not any scale associated to the 
loss of coherence at long distances and/or large times.
If we consider for example a crystal
--- let's say a semiconductor --- 
limitation to coherence in electron motion is usually produced by
thermal effects, lattice defects
and impurities present in the sample, 
which produce phase randomization of the
wavefunction components, as well as localized states,
all effects having the tendency to destroy the long-range 
coherence in the system.
At the beginning of last century,
when technology made it possible to liquefy helium, 
superconductivity was accidentally discovered
in the study of the low-temperature dependence
of electric conductivity in metals,
and a first example of a macroscopic quantum
system was found.
Nowadays, the superfluid transition at low temperatures 
of the bosonic isotope of Helium (He$_4$) and, 
at much lower temperatures, the superfluid phases 
of the fermionic one (He$_3$), 
the Bose-Einstein condensation of alkali gases, etc. 
are all well-known examples of macroscopic or mesoscopic 
quantum systems. 
In more recent times, the possibility of constructing
potential barriers on the scale of atomic
size or so by means of sophisticated growing techniques
has given rise to a new application
of quantum mechanics --- the so-called nanophysics.
%
\begin{figure}[ht]
\begin{center}
\includegraphics[width=0.5\textwidth]{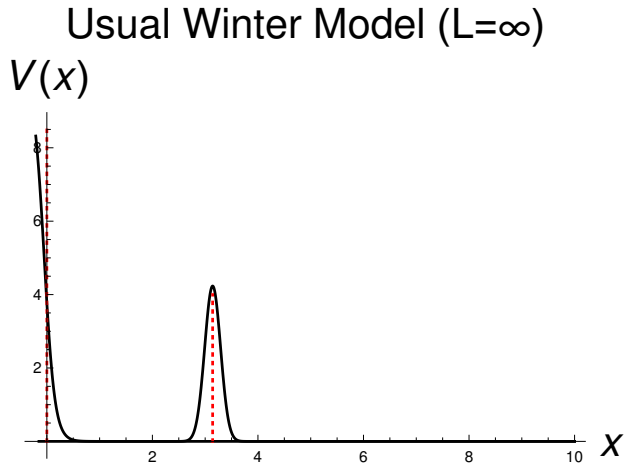}
\footnotesize
\caption{
\label{fig_pWinterinf}
\it potential energy of standard Winter model, at infinite volume 
(length $L=\infty$), or equivalently, if we work in momentum 
space, in the continuum, for a value of the coupling 
of the cavity $([0,\pi])$ to the continuum 
$([\pi,\infty))$ $g=0.2$ (see eq.$\,$\ref{H_Winter}).
For easy of visualization,
the Dirac $\delta$-potential centered at the point $x=\pi$
is approximated by a sharp Gaussian distribution 
(standard deviation $\sigma=0.15$) 
with mean value at $x=\pi$. 
The vanishing boundary condition of the wavefunction 
at the origin, $\psi(x=0;t) \equiv 0$, is represented
by a high and thick potential wall at $x<0$.
}
\end{center}
\end{figure}
%

A typical quantum phenomenon is the decay
of a resonant state --- such as a naturally unstable 
nucleus or elementary particle, or a stable particle
confined between high but slightly penetrable potential walls 
(tunnel effect).
An unstable state is a specific narrow energy
wavepacket, whose dynamics at large times is 
controlled by the interference of the 
hamiltonian eigenstates
with slightly different energies.
Because of additional wave reflection 
and interference phenomena,
even deeper coherence phenomena can be investigated in quantum 
mechanics by studying the decay of resonant states at finite,
rather than infinite, volume.
%
\begin{figure}[ht]
\begin{center}
\includegraphics[width=0.5\textwidth]{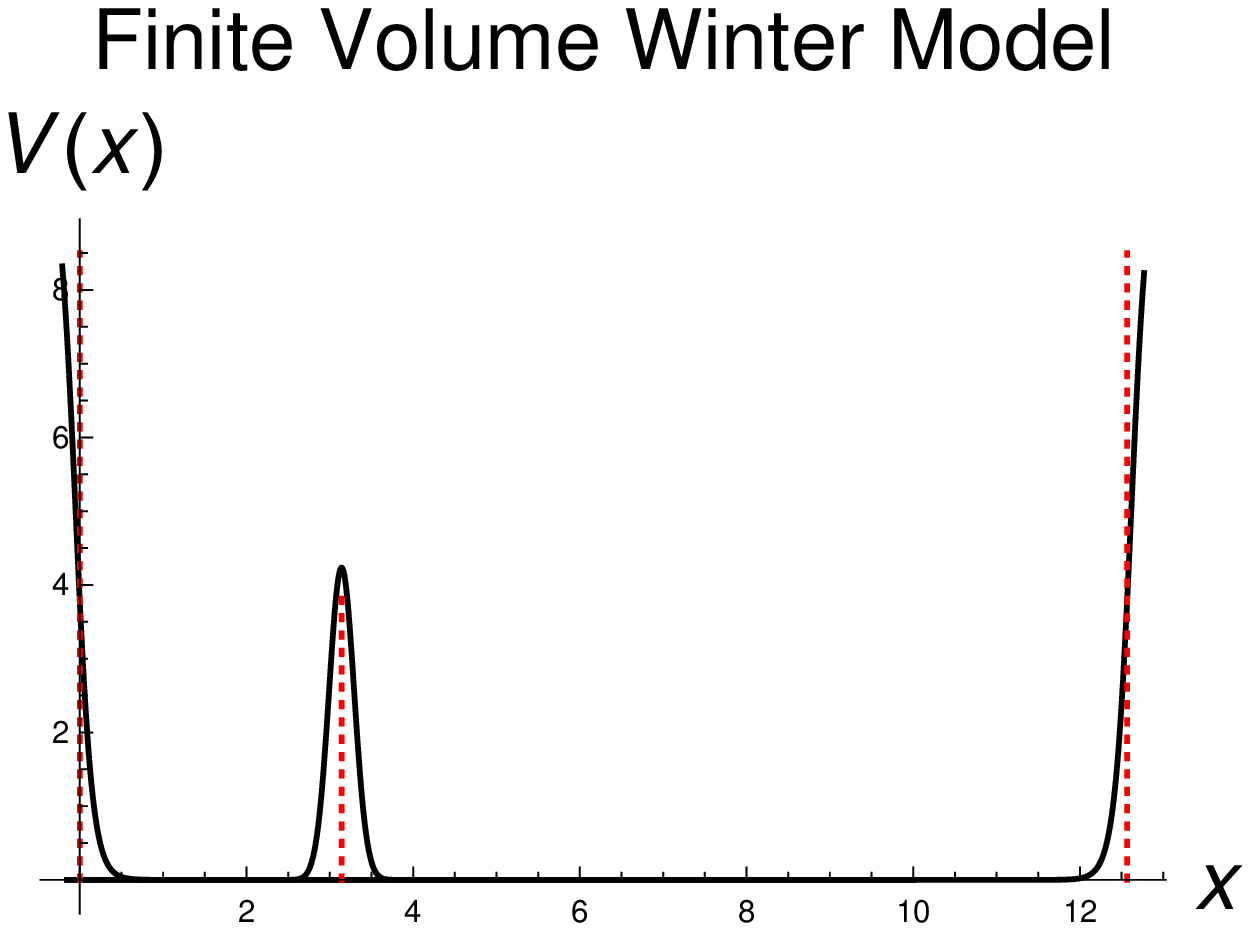}
\footnotesize
\caption{
\label{fig_pWinterfi}
\it potential energy of finite-volume Winter model 
for $L=4\pi$, i.e. for the large cavity 
($x>\pi$, to the right of the peak) three times bigger than the 
small one ($x<\pi$, on the left), 
for $g=0.2$.
For visualization purposes, as in previous plot,
the Dirac $\delta$-potential is approximated
by a sharp Gaussian distribution centered at $x=\pi$, 
while the vanishing boundary conditions of the wavefunction 
at the borders of the segment $[0,L]$, 
$\psi(x=0;t) \equiv \psi(x=L;t) \equiv 0$, 
are represented by high and thick potential walls 
at $x<0$ and at $x>L$ respectively.
}
\end{center}
\end{figure}
%
It is in this spirit that we study in this paper
resonance phenomena in the generalization
to finite volume of the so-called Winter 
(or $\delta$-shell) model.
The latter is a one-dimensional quantum-mechanical
model, describing the coupling of a resonating cavity to a 
continuum of states,
with Hamiltonian given, in proper units, by the
operator \cite{primo}  (see fig.~\ref{fig_pWinterinf})
\beq
\label{H_Winter}
\hat{H} \, = \, 
- \, \frac{\partial^2}{\partial x^2} \, + \, 
\frac{1}{\pi g } \, \delta(x-\pi);
\eeq
where $\delta(z)$ is the Dirac $\delta$-function 
centered in the origin and
$g \ne 0$ is a real coupling constant.
The motion of the particle is restricted to the positive half-line,
\beq
0 \, \le \, x \, < \, \infty, 
\eeq
and its wavefunction is assumed to vanish at the space origin
at all times%
\footnote{
Alternatively, one may think the particle
coordinate $x$ to be defined on the entire real line,
\beq
\nonumber
- \, \infty \, < \, x \, < \, \infty,
\eeq
with an infinite, impenetrable potential wall at $x<0$.
}:
\beq
\label{BC_Winter}
\psi\left( x = 0; \,t \right) \, \equiv \, 0;
\qquad t \, \in \, \RR.
\eeq
Note that standard Winter model is a one-parameter model 
--- namely the coupling $g$.

Because of the rather involved dynamics
of finite-volume Winter model (see fig.~\ref{fig_pWinterfi}),  
we limit ourselves in this work to the study
of time-independent phenomena;
we will see that the spectrum of this
model contains a considerable amount of 
information about resonance dynamics. 
Rather simple analytic formulae are obtained 
in perturbation theory up to third order in the
coupling $g$ which, 
when properly analyzed in combination with numerical 
computations, will allow us to understand some
general properties of resonance
dynamics at finite volume.
Explicit time-dependent phenomena in finite-volume
Winter model will be treated in a forthcoming publication.

It may be worth summarizing the main theoretical 
activity over the years concerning the standard (infinite-volume) 
Winter model (see fig.~\ref{fig_pWinterinf}).
As far as we know, this model was originally
introduced in ref.\cite{flugge}, 
where the resonance spectrum in the weak-coupling repulsive regime, 
$0 < g \ll 1$, was investigated, finding all the typical 
resonance phenomena \cite{newton}.
The latter can be most easily described 
in qualitative terms as follows.
In the limit $g \to 0^+$, the coefficient in front
of the Dirac $\delta$-potential on the r.h.s. of 
eq.(\ref{H_Winter}) diverges and the coupling of the 
$[0,\pi]$ cavity to the half-line $[\pi,\infty)$ 
exactly vanishes.
In this free limit, the system approaches
the union of two non-interacting subsystems:
a free particle in the box $[0,\pi]$
possessing, as well known, a discrete spectrum 
with eigenfunctions
\beq
\label{box_intro}
\phi_n(x) \, = \, \sqrt{\frac{2}{\pi}} \, \sin(nx);
\qquad 0 \, \le \, x \, \le \, \pi;
\qquad
n = 1,2,3,\cdots;
\eeq
and a free particle in the half-line $[\pi,\infty)$,
having a continuous spectrum.
If we adiabatically (i.e., in practice, very slowly) 
turn on the coupling from the free-theory value $g=0$ 
up to a very small value $g \ll 1$, 
the box eigenfunctions in eq.(\ref{box_intro}) go one-to-one 
into long-lived resonant states $(\tau \propto 1/g^2)$,
with qualitative properties similar to the former;
the only qualitative difference between the
box eigenfunctions and the resonant states is 
indeed a non vanishing width and a non-zero amplitude
at $x>\pi$ of the latter.
Winter model contains therefore an infinite,
non-degenerate resonance spectrum at small positive 
coupling.

Because of tunneling effect,
a particle initially in the $[0,\pi]$ cavity
will leave it, ending up into the infinite region 
$[\pi,\infty)$.
That implies that Winter model possesses 
in the repulsive case $g>0$, 
a continuous spectrum only; 
by explicit computation, one finds that
the model contains only a continuous spectrum 
also in the strongly-attractive case $g<-1$. 
In the weakly-repulsive case, $-1<g<0$, 
the model possesses also a discrete spectrum,
containing a single bound state, with the particle
exponentially localized close to the negative
barrier at $x=\pi$ \cite{primo}.

R.$\,$Winter in his well-known paper \cite{winter} 
(see also \cite{gottfried} and \cite{dicus})
studied the time evolution of wavefunctions 
$\phi(x;t)$ equal, at the initial time $t=0$, 
to the box eigenfunctions in eq.(\ref{box_intro})
--- particle initially inside the cavity.
He used both perturbative and numerical
methods.
An asymptotic power decay for $t \to + \infty$ was 
explicitly demonstrated for the first time.
In ref.\cite{primo} all the calculations originally
made in ref.\cite{winter} were checked, finding complete 
agreement as far as the exponential decay of the first 
resonance is concerned, as well as the asymptotic power 
decay of all the resonances. 
A disagreement was found instead in the exponential
decay of the excited resonances ($n>1$) because of 
the occurrence of a peculiar state-mixing phenomenon.
Each initial box eigenfunction (eq.(\ref{box_intro})),
labeled by the quantum number $n$,
does not couple only to the corresponding resonance,
of the same order $n$, but actually to all the resonances.
The diagonal couplings are 
$1 + \mathcal{O}(g)$, i.e. are large,
while the off-diagonal couplings are 
$\mathcal{O}(g)$, i.e. small ($|g| \ll 1$),
but may give rise to leading effects in large temporal 
regions.
If we consider for example the evolution of an unstable
state equal at $t=0$ to the box eigenfunction 
in eq.(\ref{box_intro}) with $n=2$,
it is found that in a large (sub-asymptotic) temporal window, 
the dominant contribution to this state comes from the 
first resonance $(n=1)$, rather than from the second one 
$(n=2)$.
That is because the contribution from
the fundamental resonance, 
\beq
\approx \, g \, \exp\left[ - \, \Gamma_1(g) \, t \right],
\eeq
even though involving a small
$\mathcal{O}(g)$ coefficient, 
has a slower exponential time decay
than the contribution coming from the first excited resonance,
\beq
\approx \, \exp\left[ - \, \Gamma_2(g) \, t \right],
\eeq
which has a large $\mathcal{O}(1)$ coefficient,
but decays much faster. 
The decay rates $\Gamma_n(g)$ grow as $n^3$ (see later),
so that
\beq
\Gamma_2(g) \, \simeq \, 8 \, \Gamma_1(g).
\eeq
The smaller decay width $\Gamma_1(g)$ 
"wins" the smaller expansion coefficient $\mathcal{O}(g)$%
\footnote{
Qualitatively similar phenomena occur
in the numerical computations of two-point
correlation functions in lattice QCD.
By neglecting multi-particle continua,
an interpolating operator $O(x)$, acting on the vacuum,
creates a superposition of hadronic states
with different masses $m_n$ and different strengths $Z_n$,
\beq
\int d^3 x \, \langle 0 \left| \, O(x) \, O(0) \, \right| 0 \rangle 
\, \simeq \, \sum_{n=1}^\infty \frac{Z_n}{2m_n} \, e^{- m_n t}.
\eeq
In this case, the masses rather than the rates,
control the exponential decay with time.
}.
At first order in $g$, the mixing of states
is controlled by an infinite real antisymmetric 
matrix, which was explicitly calculated in ref.\cite{primo}.

In ref.\cite{secondo} the idea 
that the resonance wavefunctions can be obtained
from the simple box wavefunctions (eq.(\ref{box_intro}))
by means of finite renormalization of the parameters
entering the latter,
was verified by means of higher-order perturbative 
computations.
The parameters in question are the momentum 
of the particle $k_n(g) = n + \mathcal{O}(g)$,
which receives a real correction at first order in $g$
and both a real and an imaginary correction at second order,  
its energy $E_n = k_n^2$, plus an overall normalization
constant $Z_n(g) = 1 + \mathcal{O}(g)$.
Actually, that is nothing but the formalization of the physical 
idea that resonance states of the Winter model have to be close to  
the box eigenfunctions (i.e. are "perturbations" of the latter) 
for sufficiently small couplings.
Since any special orthogonal (rotation) matrix is, to first order 
in the angles, a real antisymmetric matrix, one may guess 
that the state-mixing matrix of Winter model will to be equal,
after the inclusion of the higher order terms, to an infinite 
orthogonal matrix.
By explicitly computing such a matrix to second order 
in $g$, it was found  that, in addition 
to the expected iteration of the first-order matrix, 
there do also exist additional contributions whose interpretation
in the above sense is problematic and which
do not seem consistent with such an idea.

In order to understand the limitations
of perturbation theory in the analysis of 
Winter model, the latter was studied in 
ref.\cite{lungo} with geometrical, rather 
than analytical, methods.
By complexifying the coupling $g$, namely 
\beq
g \, \in \, \RR 
\quad \mapsto \quad 
g \, \in \, \CC, 
\eeq
it was found that each resonance,
anti-resonance, bound and anti-bound state
of the model is related to a specific 
sheet of the Riemann surface of a basic 
transcendental infinite-valued function.
In particular, by knowing the singularity
structure of such surface,
the convergence radii $R_n$'s
of the expansions of the resonance momenta $k_n$
in powers of $g$, $k_n = k_n(g)$, were determined;
they turned out to be all strictly positive,
for each $n=1,2,3,\cdots$, but going to zero
as $1/n$ for $n \to \infty$.
For low $n$ (the cases of practical 
interest), the $R_n$'s were found to be actually 
one order of magnitude smaller than expected.
The problem of constructing initial
wavefunctions exciting exactly one resonance at a time,
rather than a superposition, was also considered 
in ref.\cite{lungo};
formally one has to invert the state-mixing matrix.
Only an approximate solution of this problem
was found, as well known from numerical
studies ("you never fill a resonant state" \cite{nic,nic2}).

As already discussed, in the weakly-attractive case, 
$-1 \ll g < 0$, Winter model contains a bound state $(bs)$
having an energy 
\beq
E_{bs} \, \approx \, \frac{1}{g} \, \to \, - \, \infty 
\qquad \mathrm{for} \,\,\, 
g \, \to \, 0^-.
\eeq
The spectrum of Winter model is therefore unbounded from below 
in any (complete) neighborhood of the
free-theory point $g=0$, and general instabilities
phenomena may occur.
Similar instabilities are believed to occur also in 
Quantum-Electro-Dynamics (QED) for any negative 
value of the fine structure constant, $\alpha<0$.
Since an exact solution of interacting quantum
field theories such as QED is beyond human abilities 
\cite{Aglietti:2018aqc},
a motivation for the study of ref.\cite{lungo}
was also to understand, by means of the above analogy, 
some properties of the vacuum structure in quantum field theory.

In ref.\cite{Aglietti:2016zzm} a general expansion 
in powers of $1/n$ for highly-excited resonance states 
was introduced, where $n \gg 1$ is the excitation number
($n=1$ for the fundamental resonance).
This expansion allows an approximate resummation
of the perturbative series to all 
orders in $g$ for the decay rate of the resonances,
as well as for other observables.
In the case of the Winter model, the $1/n$ expansion 
gives for the decay rate, in Leading Order ($LO$), 
\beq
\label{Gamma_resummed}
\Gamma_n^{(LO)}(g) \, = \, 
\frac{n}{\pi} \log\left[ 1 + (2\pi n g)^2 \right].  
\eeq
The above formula resums all the terms of the
form
\beq
g^2 n^3, \,\, g^3 n^4, \,\, g^4 n^5, \,\, \cdots,
\eeq
i.e. with the  power of $n$ minus the
power of $g$ equal to one:
$g^h n^{h+1}$, $h=1,2,3,\cdots$.
By expanding the logarithm on the r.h.s. of 
eq.(\ref{Gamma_resummed}) in powers of $g$,
one recovers the well-known lowest-order ($lo$) 
perturbative formula \cite{primo,winter}
\beq
\Gamma_n^{(lo)}(g) \, = \, 4 \pi \, n^3 g^2;
\qquad n \, g \, \ll \, 1;
\eeq
while, for highly-excited resonances,
the asymptotic $(as)$ formula is obtained \cite{lungo}
\beq
\Gamma_n^{(as)}(g) \, \simeq \, 
\frac{2}{\pi} \, n \log\left(n g \right);
\qquad n \, \gg \, \frac{1}{g} \, \gg \, 1.
\eeq
Note that, as an effect of the resummation,
the leading-order $n^3$ growth of the width
$\Gamma_n(g)$ with the resonance-order $n$
is converted into a much softer $n \log n$
behavior.

In refs.\cite{nic} and \cite{delaMadrid:2015ooa}
Winter model
was used as a check of general formalisms
for unstable states.
Even though this model is,
because of the its schematizations
and its low-dimensionality,
mostly a mathematical-physics model,
it has been used in ref.\cite{segre}
for a semi-quantitative description
of $\alpha$ decay in heavy nuclei.
A review of generalizations of Winter model
and applications of the latter
to quantum chemistry can be found in the Introduction 
of ref.\cite{delaMadrid:2017oeo}.

In order to make the paper reasonably self-contained
and therefore accessible also to non-experts of the field,
we have written a few review sections 
before the ones containing original results
(the experienced reader may begin to read the article
from sec.$\,$\ref{wint_mod_fin_vol}).
The paper is organized as follows.

In sec.$\,$\ref{res_inf_vol} we summarize
the standard theory of resonances both in 
classical mechanics and in quantum mechanics.
According to the latter, the well-known
exponential decay law of unstable particles
only holds approximately in an intermediate 
time region;
the pre-exponential and the post-exponential
regions are briefly described.

In sec.$\,$\ref{res_fin_vol} we describe 
in qualitative terms
the new features of resonance dynamics which occur 
when we go from infinite volume to finite (large) volume;
in momentum space, that means to go from
a continuous spectrum to a quasi-continuous spectrum
of the decay channels.
A typical phenomenon which occurs at finite volume 
is recursion in time.
Given the (in principle unlimited) coherence of 
quantum mechanics, with some fantasy 
we can imagine that one day given, let's say, 
the decay of a $Z^0$ into a muon pair,
\beq
Z^0 \, \to \, \mu^+ \, + \, \mu^-,
\eeq
it will be possible, by means of some
experimental apparatus, to make the decay products
collide with each other to produce back a $Z^0$,
\beq
\mu^+ \, + \, \mu^- \, \to \, Z^0.
\eeq	
This "$Z^0$ regeneration phenomenon" from
a given decay channel provides a high-energy, 
"extreme" example of a recurrence phenomenon
in resonance dynamics.
Recursion is actually a general phenomenon
occurring in autonomous quantum systems 
$(\partial \hat{H}/\partial t = 0)$ 
possessing discrete spectrum only.
It is the quantum analog of the classical
recurrence described by the Poincare' theorem.
A specific phenomenon of resonances at finite volume
is the so-called limited quantum decay
--- namely the fact that, unlike at infinite volume,
not all the particles in the sample 
eventually decay. 

In sec.$\,$\ref{stand_wint_mod}
we summarize the main features of standard
Winter model (infinite volume) which, as discussed above,
describes a particle confined in the segment $[0,\pi]$,
between an impermeable wall (at $x=0$) 
and a slightly permeable one (at $x=\pi$).
This simple model allows for an elementary analytic 
description of many typical resonant phenomena.
Because of the reflecting boundary condition at
the space origin (eq.(\ref{BC_Winter})), 
the momenta $k$ of the particle are only defined 
up to a sign, so that we can assume for example $k>0$;
that implies that the energy spectrum 
of standard Winter model is not degenerate.

In sec.$\,$\ref{wint_mod_fin_vol} we discuss 
the generalization of Winter model 
to finite volume (see fig.~\ref{fig_pWinterfi}).
The latter describes a small resonating cavity
weakly coupled to a large one;
the usual model is recovered in the limit in which 
the length of the large cavity is sent to infinity.
In this model, recursion in time is the fact that,
if we put the particle initially inside the small cavity,
$x \in [0,\pi]$, it will leave the latter 
after some characteristic time,
but it will go back inside it with a probability arbitrarily 
close to one after a sufficiently long, but finite,
evolution time.

In sec.$\,$\ref{spec_vol_fin} we compute the
spectrum of finite-volume Winter model.
Actually, for technical reasons,
we consider the momentum spectrum
rather than the energy spectrum ($E = k^2$).
The momentum spectrum of the particle contains 
exactly-integer momenta, 
\beq
p_n \, = \, n = 1, 2, 3, \cdots,
\eeq
for any value of the coupling $g$,
which can be considered "exceptional".
Corresponding to the exceptional momenta, 
there are "exceptional eigenfunctions",  
which also do not depend on $g$ and have exactly 
equal amplitudes inside both cavities.
There are also non-integer, ordinary, momenta,
which are the solutions of a real transcendental equation
containing the coupling $g$ (cfr. eq.(\ref{eq_basic})).
The normal eigenfunctions, corresponding to the non-integer 
momenta, depend explicitly on $g$;
some of them have a resonating behavior, i.e.
have an amplitude inside the small cavity (much) larger
than inside the large one, or an anti-resonating
behavior, i.e. they have a small-cavity amplitude
smaller than the large-cavity one.
The exceptional eigenfunctions, in some sense,
"separate" the resonating eigenfunctions from
the anti-resonating ones. 

Since the transcendental equation satisfied
by the ordinary momenta cannot be solved
in closed analytic form,
we derive in sec.$\,$\ref{ord_pert_th}
ordinary perturbative expansions
for the allowed momenta $k=k(g)$
of the particle in powers
of the coupling $g$.
From third order in the coupling $g$ on, secular terms of the
form 
\beq
\label{secular}
g^3 N, \quad g^4 N^2, \quad g^5 N^3, \quad g^6 N^4, \cdots
\eeq
are found in the perturbative expansion of the momenta
(some coefficients may actually vanish).
The parameter $N$ is the ratio of the length
of the large cavity over the small one
and is assumed to be integer, as the model
is simpler in this case.
The terms (\ref{secular}) tend to spoil the convergence of the 
perturbative expansion in the large-volume case $N \gg 1$.
The control over such terms is in any case necessary 
to deal with the infinite-volume limit $N \to \infty$.

In sec.$\,$\ref{res_pert_th} we derive improved
perturbative expansions for the quantized momenta
$k=k(g)$, which approximately resum the secular
terms above to all orders in $g$.
By comparing ordinary and improved
perturbative expansions of the momenta $k = k(g)$,
with exact (numerical) computations of the latter
in a large region of $g$,
we find that the resummed formulae 
describe the spectrum much more accurately than the 
fixed-order ones.

Finally, in sec.$\,$\ref{sec_concl} we draw the 
conclusions of our first analysis of Winter
model at finite volume.
We also discuss how such system can be 
physically realized and which phenomenology
is expected.


\section{Resonance Dynamics at Infinite Volume}
\label{res_inf_vol}

In this section we review some general properties
of resonance decay at infinite volume
both in classical and quantum mechanics;
more details can be found in the review \cite{Fonda:1978dk}.


\subsection{Classical Mechanics}

Within classical mechanics, it is not possible to describe
the internal dynamics of any unstable state such as, for example, 
an unstable nucleus;
one can only derive the well-known exponential decay
law with time as follows.
Given a sample of $N\gg 1$ identical nuclei,
it is natural to assume that the number $dN$ of decaying 
ones in the elementary time interval $dt$ is proportional
to $dt$ itself;
if we further assume, as it is natural to do, 
that the nuclei do not significantly interact
with each other, it follows that $dN$ is also
proportional to the number $N=N(t)$ of undecayed nuclei 
present in the sample at the current time $t$
so that, at the end:
\beq
\label{decay_class}
dN \, = \, - \, \Gamma \, N(t) \, dt,
\eeq
where $\Gamma>0$ is a proportionality 
constant dependent on the nuclei species and, 
in general, also on time.
However, if time $t$ is homogeneous
 --- i.e. physics is invariant under arbitrary time translations --- 
then $\Gamma$ cannot depend on time
and is therefore a constant characterizing 
the nuclei under study.
By integrating eq.(\ref{decay_class}) with respect to time, 
one immediately obtains the well-known exponential decay law: 
\beq
N(t) \, = \, N(0) \, e^{ - t/\tau } ;
\eeq
where
\beq
\tau \, \equiv \, \frac{1}{\Gamma}
\eeq
is the mean lifetime of the nuclei under study.
The decay width $\Gamma$ enters the above equation simply as a proportionality constant, 
so it cannot be calculated/predicted in classical mechanics.


\subsection{Quantum Mechanics}

In quantum mechanics, we can say that, in general,
resonances occur when a discrete state is immersed and 
weakly-coupled to a continuum
\cite{newton, nic, nic2, Fonda:1978dk, gamow, mt}.
When expressed as superposition
of energy eigenstates 
of the real (interacting) system, 
a resonance is represented by a 
narrow wave packet and is therefore a 
quasi-stationary state --- a kind of unstable state.
As discussed in the Introduction,
quantum coherence plays a fundamental role in resonance dynamics;
indeed, to completely destroy a resonant state, 
it would be sufficient to change 
the phase of a single (not too small) coefficient 
entering the spectral decomposition
of its wave function.

A fundamental quantity in the study of unstable systems
is the so-called No-Decay (ND) probability, defined as
\beq
P_{ND}(t) \, \equiv \, \left| A_{ND}(t) \right|^2,
\eeq
where $A_{ND}(t)$ is the No-Decay amplitude,
\beq
\label{A_ND}
A_{ND}(t) \equiv \left\langle \psi(0) | \psi(t) \right\rangle
\, \equiv \, 
\left\langle \psi(0) \Big| \hat{U}(t;0) \Big| \psi(0) \right\rangle,
\eeq
with $U\left(t_2;t_1\right)$ the Schrodinger time-evolution
operator,
\beq
\hat{U}\left( t_2; t_1 \right) 
\, \equiv \, e^{-i \hat{H} \left(t_2-t_1\right)};
\qquad
t_1,\, t_2 \, \in \, \RR;
\eeq
and $\hat{H}$ the Hamiltonian operator of the system
assumed, for simplicity's sake, to be time-independent
($\partial \hat{H}/\partial t \equiv 0$).
The physical interpretation of eq.(\ref{A_ND}) is 
straightforward:
given the unstable state $\left| \psi(0) \right>$,
prepared at the initial time $t=0$, 
in order to determine the no-decay amplitude,
we just project $\left| \psi(t) \right>$, 
the evolved state from $t=0$ to the generic time $t$, 
on the original, unevolved state $\left| \psi(0) \right>$.
Exponential decay which, as we have seen above,
is an exact law in classical physics, valid at all times,  
is not in general a straightforward
and exact consequence of the Schrodinger evolution
\cite{Torrontegui}.
The point is that Schrodinger equation is a dispersive 
wave equation, which naturally describes oscillations in space and time, 
quite different from, let's say, the heat equation, 
generating "kinematically" exponential time decays.
One can usually identify three different
temporal regions in the decay of an unstable system
in quantum mechanics, which are schematically
discussed in the next sections.


\subsubsection{Small-time region}

Since the first derivative of the no-decay probability
can be shown to vanish at the initial time $t=0$ 
--- the so-called Zeno effect \cite{Fonda:1978dk} ---
\beq
\frac{dP_{ND}}{dt}(t=0) \, = \, 0,
\eeq
it follows that the latter has a quadratic dependence
on time at small times,
\beq
\label{small_time}
P_{ND}(t) \, = \, 1 \, - \, c \, t^2 
\, + \, \mathcal{O}\left(t^3\right),
\eeq
where $c$ is a positive constant.
Since a linear term in $t$ is absent in the above
equation, the particle is "late at decay"
and its decay is not exponential in this region.
Unlike the intermediate and asymptotic time regions (see later), 
the small-time region is strongly dependent on the details 
of the preparation of the initial state.
In physical terms, we may say that the system,
once prepared at the initial time $t=0$,
"adjustes" its initial wavefunction for beginning 
to decay \cite{winter}.
The occurrence of a time behavior of the form (\ref{small_time})
has been experimentally verified.


\subsubsection{Exponential region}

This is the intermediate region where the well-known
exponential decay with time occurs to a 
(very) good approximation:
\beq
\label{exp_plus_Z}
P_{ND}(t) \, \simeq \, Z \, e^{-\Gamma t};
\eeq
where $Z$ is a normalization constant
$(0 < Z	\le 1)$ \cite{primo,secondo,lungo}.
The exponential behavior above which, 
as discussed above, holds exactly in classical physics,
is true only approximately in quantum mechanics and 
in a limited time interval.
The law described by eq.(\ref{exp_plus_Z}) is the 
one usually found in experiments, in the sense
that experimental data turn out to be well fitted 
by an equation of this form up to the largest times
investigated.
E.$\,$Rutherford was the first one to check the above law 
in $\alpha$-decay of heavy nuclei for twenty 
lifetimes or so \cite{nic}.


\subsubsection{Asymptotic region}

Finally, it does also exist an asymptotic region in which
the no-decay probability is proportional to an inverse 
power of time:
\beq
\label{large_times}
P_{ND}(t) \, \simeq \, \frac{k}{t^p};
\qquad t \gg 1;
\eeq 
with $k$ is a positive constant and $p$
is an integer index of order one.
In the Winter model, for example, $p=3$ \cite{primo,winter,secondo,lungo}.
In general, interference effects occur in the transition
interval from the exponential region to the power, 
post-exponential one,
where these two contributions to the no-decay
probability are of comparable size. 

It can be shown by means of a general
argument based on analyticity properties of the Fourier Transform, 
that $P_{ND}(t)$ cannot decay exponentially for 
$t \to + \infty$ but has to decay, roughly speaking, slower than 
that \cite{khalfin}.
In physical terms, we may say that the unstable particles
in the sample under study decay, at very large times, 
less frequently than expected according to
the classical reasoning.
In practice, in the post-exponential region,
a power behavior with time is always found, 
with different values of the index $p$ 
in the different models analyzed so far
(dependent on the spin of the resonance,
threshold effects, etc. \cite{mt}).
However, let us remark that,
in any case, time evolution
completely empties the initial state
asymptotically \cite{Fonda:1978dk}:
\beq
\label{complete_decay}
\qquad
\lim_{t\to+\infty} P_{ND}(t) \, = \, 0
\qquad
\mathrm{(Infinite\,\,Volume).}
\eeq 
In other words, by waiting for a long enough time,
all the unstable particles in the sample
under observation are eventually found to decay.
If the no-decay probability is written
as a Fourier transform over the energy,
the total decay at asymptotic times
given by eq.(\ref{complete_decay}) 
is just a consequence of the well-known
Riemann-Lebesgue lemma of Fourier Transform
Theory. 
Furthermore, apart from very small interference effects \cite{winter},
the no-decay probability $P_{ND} = P_{ND}(t)$
is generally a strictly monotonically
decreasing function of $t$.

The experimental observation of 
post-exponential effects
as described by eq.(\ref{large_times})
is traditionally very problematic --- an obvious
reason for that being that at the very large times
needed for that, the signal-to-background ratio 
becomes quite small.
It has been claimed that a decay law of the form
(\ref{large_times}) is observed in some organic molecules, 
but the theory is not elementary in this case
and interpretation problems come into play \cite{nic}.
It has also been speculated that a power behavior
of the form (\ref{large_times})
never sets in in systems where repeated measurements
at different times are performed because of a quantum
"resetting" effect \cite{Fonda:1978dk}.


\section{Resonance Dynamics at Finite Volume}
\label{res_fin_vol}

In this section we describe some general phenomena
which occur in resonance decay at finite 
--- rather than infinite --- volume;
if we work in momentum space rather than in physical 
(configuration) space, by going to finite volume, 
the continuum states of the decay products 
of the unstable state are replaced by a set of 
corresponding discrete levels.
If the volume is large, i.e. we are close to
the continuum, the levels are closely-spaced.
The simplest system consists of a 
discrete state immersed and weakly coupled to a 
Quasi-Continuum (Q.C.) of states, rather than to a 
Continuum (C.) of states, as in the usual,
infinite-volume, case \cite{fano}.
%
\begin{figure}[ht]
\begin{center}
\includegraphics[width=0.5\textwidth]{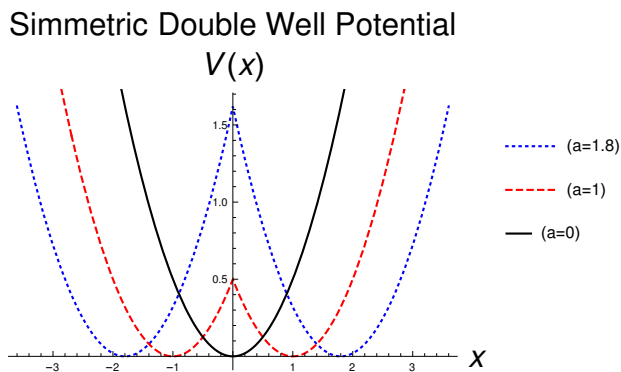}
\footnotesize
\caption{
\label{fig_pmerzbacher}
\it plots of the symmetric double well potential
$V(x) = \omega^2 \big(|x|-a\big)^2/2$ for $\omega=1$
and different values of the parameter $a$: 
$a=0$, where the potential reduces to a quadratic one 
(black continuous line);
$a=1$ (red dashed line) and 
$a=1.8$ (blue dotted line).
}
\end{center}
\end{figure}
%

The general problem of resonance dynamics at finite volume
is also of interest in lattice quantum field theory
--- namely lattice QCD ---
where the quantum fields are regularized by introducing
a finite euclidean space-time lattice \cite{libri_ret}%
\footnote{
The euclidean time is just an additional space
coordinate.
}.
The theory so constructed possesses both an ultraviolet 
energy/momentum cutoff $\Lambda_{UV}$, of the order of 
the inverse of the lattice spacing $a$,
\beq
\Lambda_{UV} \, \approx \, \frac{1}{a};
\qquad
a \, > \, 0;
\eeq
as well as an infrared cutoff $\lambda_{IR}$, given by the inverse
of the lattice length $L$, 
\beq
\lambda_{IR} \, \approx \, \frac{1}{L};
\qquad
L \, < \, \infty.
\eeq
The two length scales above are related by the equation
\beq
L \, = \, N a, 
\eeq
where $N\gg 1$ is the number of lattice points along one of the 
coordinate axes (the lattice size). 
The unavoidable finite-size effects,
$V<\infty$ or, equivalently, 
in energy-momentum space, $\lambda_{IR}>0$,
are currently used to determine scattering quantities 
\cite{Luscher:1985dn}.
The main application so far developed is the calculation
of $g_{\rho\pi\pi}$, the $\rho\pi\pi$ coupling entering the
dominant, strong-interaction decay
of the rho meson into two pions,
\beq
\rho \, \to \, \pi \, + \, \pi.
\eeq
In order to understand in a simple way the 
relevant physics
at finite volume, let us consider 
the scattering of a non-relativistic particle on a 
short-range repulsive potential $U=U(r)$.
In the usual case of an infinite volume available
to the particle (typically the whole space $\RR^3$), 
the initial state of the particle is described, 
in the far past $t \approx -T$ $(T \gg 1)$, 
by a wavepacket far away from the potential range,
with some average momentum $\bar{\bf{p}}$.
When the wavepacket hits the potential $U(r)$, 
i.e. reaches the potential region
(where the potential is significantly different from zero),
let's say at $t \approx 0$, 
an outgoing wave having the form of a spherical shell
of thickness of the order of the packet size is generated 
by the interaction.
By progressively increasing the time $t$ from 
the interaction time $t \approx 0$
up to the far future, $t \approx T$, one ends up with the
the original wavepacket evolved from $-T$ to $+T$
(which has simply traveled with the average velocity 
$\mathbf{v} = \mathbf{p}/m$), 
superimposed to a spherical shell
emerging out of the space origin $r=0$ from $t \approx 0$  
with the radial velocity $v_r = \left| \bf{v} \right|$.
When we go to finite volume, the energy spectrum 
of the system changes from a continuous to
a discrete one.
Furthermore, since the particle is now always at a finite distance
from the potential $U(r)$, the energy levels are modified 
with respect to the free case $U(r)\equiv 0$;
according to physical intuition, we expect such a level shift 
to increase by lowering the volume.
While in the continuum-spectrum case (infinite volume), 
the potential causes transitions ($S$-matrix), 
in the Quasi-Continuum case  (finite volume), $U(r)$ causes 
calculable shifts of the discrete energy levels.
By looking at the spectrum of the system at different volumes, 
it is possible to reconstruct the scattering phases
of the infinite-volume system.

In general, as one could expect according to
physical intuition, by going from infinite volume to finite volume,
resonance dynamics becomes far more complicated
and new phenomena, which do not occur at infinite volume,
come into play.
The main ones are: 1. recurrence and 2. limited decay.
Let us briefly discuss them in the following sections.


\subsection{Time Recurrence}
\label{sec_recur}

Time Recurrence in a quantum system 
can be roughly defined as the fact that the latter goes back
to its initial state with arbitrary (high) degree of accuracy
after a proper, sufficiently long but finite time evolution.
Recurrence was originally discovered in classical mechanics,
where the following theorem holds.


\subsubsection{Poincare' Theorem}

In classical mechanics, the following recurrence
theorem holds \cite{vulpiani}:

\vskip 0.3truecm
\noindent
{\bf Th.:} {\it given an autonomous Hamiltonian system, 
\beq
\frac{\partial H}{\partial t} \, \equiv \, 0, 
\eeq
with a finite number of degree of freedom,
having a finite-volume phase space
and given an arbitrary time $T>0$,
an arbitrary (small) neighborhood
of any initial state ($t=0$) 
intersects itself under the Hamiltonian flow at some 
time $t>T$.
}

\vskip 0.3truecm
\noindent
In practice, a finite number of degrees of freedom
means a finite number of particles and the hypothesis 
of finite volume of the phase space 
is verified for example if the particles  of the system 
are confined into a finite region 
(let's say a box) 
and the inter-particle potentials are bounded from below
(they describe, for example, repulsive forces).
In physical terms, the system goes back arbitrary close 
to basically all its previously reached states, 
after a sufficiently long evolution time.
Basic questions related to this phenomenon
concern the dependence
of the distribution of the recursion times 
$\left\{t_i\right\}$ on the number of 
degrees of freedom in the system
and on the kind of interactions.
In general, the recursion times grow
exponentially with the number of degrees
of freedom of the system \cite{vulpiani}.
It would also be interesting to know
the size of the time slices 
$\left\{ \delta t_i \right\}$ 
where recurrence holds.


\subsubsection{Quantum Recurrence Theorem}

Even though phase space does not exist 
in quantum mechanics, an analog theorem 
to the Poincare' one holds true \cite{bocchieri}:

\vskip 0.3truecm
\noindent
{\bf Th.:}
{\it if a quantum system, described by an Hamiltonian
operator $\hat{H}$, is autonomous,
\beq
\frac{ \partial \hat{H} }{\partial t} \, = \, 0,
\eeq
and possess a discrete spectrum only, 
then its comes back to anyone of its previously-reached 
states, in quadratic norm, as close as we desire,
after a sufficiently long evolution time $t$
larger than any given threshold $T>0$.
}

\vskip 0.3 truecm
\noindent
The following remarks are in order.
\begin{enumerate}
\item
The hypothesis of time independence of the
Hamiltonian function/operator is the same 
in the classical and in the quantum cases.
\item
The assumption, in the quantum case, of discrete spectrum only 
is the natural translation of the condition of finite volume 
of phase space of the classical system.
\end{enumerate}
In order to understand the physical meaning of this theorem,
let us sketch its proof, which is based on the theory
of almost-periodic functions \cite{haraldbohr}
\footnote{
Not to be confused with the quasi-periodic functions.
}.
By expanding a general wavefunction $\phi$ 
of the system under consideration in energy 
eigenstates $\left\{ \psi_n \right\}$, 
one obtains:
\beq
\label{phi_expand}
\phi(x,t) \, = \, \sum_{n=1}^\infty
a_n \, \psi_n(x) \, e^{-i\omega_n t};
\eeq
where the coefficients $a_n \in \CC$ 
satisfy the completeness relation 
\beq
\label{completeness}
\sum_{n=1}^\infty \left| a_n \right|^2 \, = \, 1.
\eeq
We have assumed that the initial wavefunction
and the eigenfunctions of the system 
are normalized to one:
\beq
\| \phi \| \, = \, 1; 
\qquad
\left\| \psi_n \right\| \, = \, 1; 
\qquad 
n = 1, 2, 3, \cdots.
\eeq
Without any generality loss, we can assume
the frequencies (energies) of the system
to be ordered
\beq
\qquad\qquad
\omega_n \, \le \, \omega_{n+1};
\qquad
n = 1,2,3,\cdots.
\eeq
As far as the time variable $t$ is concerned,
the expansion on the r.h.s. of eq.(\ref{phi_expand}) 
is a Generalized Fourier Series,
the latter being defined by a formal trigonometric series 
of the form
\beq
\label{Four_gen}
f(t) \, \sim \, \sum_{n=1}^\infty c_n \, e^{-i\omega_n t},
\eeq
with the $c_n$'s arbitrary (complex) coefficients.
Since the frequencies $\left\{\omega_n\right\}$ have generically
irrational ratios, 
\beq
\frac{\omega_m}{\omega_n} \, \notin \, \QQ;
\qquad
m \, \ne \, n;
\eeq
the function $f=f(t)$ is not a periodic function 
of the time $t$.
\footnote{
Ordinary Fourier Series are obtained as the
particular case of eq.(\ref{Four_gen}) 
in which all the frequencies $\omega_n$
are exact integer multiples of the fundamental
frequencies $\omega_1$:
\beq
\qquad\qquad
\omega_n \, = \, n \, \omega_1. 
\eeq
In this "exceptional" case 
--- as well known from Fourier series theory ---
the function $f(t)$
is (exactly) periodic with respect to time $t$
with fundamental period $T_1 \equiv 2\pi/\omega_1$:
\beq
f\left(t+T_1\right) \, = \, f(t);
\qquad t \, \in \, \RR.
\eeq
}
%
Since, according to eq.(\ref{completeness}),
the series on the r.h.s. of eq.(\ref{phi_expand})
is convergent (in quadratic norm), we can approximate the
wavefunction $\phi(x,t)$, as accurately as we wish, 
by neglecting sufficiently small coefficients;
the function series on the r.h.s. of eq.(\ref{phi_expand})
then becomes a finite (possibly big) sum:
\beq
\phi(x,t) \, \simeq \, \sum_{k=1}^N
a_{n_k} \, \psi_{n_k}(x) \, e^{-i\omega_{n_k} t};
\eeq
As well known in the theory of almost-periodic functions,
the frequencies occurring in such a finite sum,
\beq
\omega_{n_1}, \, \omega_{n_2}, \, \cdots, \omega_{n_N},
\eeq
can be simultaneously approximated, with arbitrary accuracy, 
by means of a sequence of rational numbers with a small, 
common denominator $M$, 
\beq
\omega_{n_k} \, \simeq \, 
\Omega_k \, \equiv \, \frac{m_k}{M} \, \in \, \QQ;
\qquad 
m_k, \, M \, \in \, \NN;
\qquad
k = 1, 2, \cdots, N.
\eeq
In this way we obtain an (exactly) periodic function of time
with period $T=2\pi/M$,
which provides the required approximation:
\beq
\phi(x,t) 
\,\, \approx \,\,
\varphi(x,t) 
\, \equiv \, 
\sum_{k=1}^N a_{n_k} \, \psi_{n_k}(x) \, 
\exp\left( - i \, \frac{m_k}{M} \, t \right).
\eeq
As in the classical case, also in the quantum case
it is  interesting, once fixed a degree of accuracy (error), 
to determine the recurrence times 
$\left\{ t_i \right\}$ of the states of the system,
together with the size of the time slices 
$\left\{ \delta t_i \right\}$ in which the 
wavefunction is close to the initial one.


\subsubsection{Examples}

In ref.\cite{milonni} resonance dynamics
in the quasi-continuum limit
was studied in abstract form by considering
a system of linear first-order ordinary
differential equations (ode's) in time, describing the
evolution of the amplitude $a=a(t)$ of a discrete state
coupled to quasi-continuum amplitudes $b_i=b_i(t)$,
$i=1,2,\cdots,N$ $(N \gg 1)$.
By assuming a simple form of the coupling terms 
and, as initial condition, only the discrete state 
to be populated at $t=0$,
\beq
a(t=0) \, = \, 1; \qquad b_i(t=0) \, = \, 0;
\qquad 
i=1,2,\cdots,N
\eeq
the model was exactly solved in terms
of Laguerre polynomials and
recurrence phenomena were explicitly found.
In this particular case, the recurrence times
can easily be determined.


\subsection{Limited Decay}
\label{lim_decay}

The second new phenomenon which occurs 
in resonance dynamics when we go to finite
volume, is the fact that a finite fraction 
of unstable particles in the initial sample  
may not decay at any time.
As discussed above, this "failure to decay"
cannot occur in the continuum case (infinite
volume).
Let us remark that this phenomenon is
not as general as the time recurrence
which we have considered in the previous section;
in particular, there is not any general theorem
concerning limited decay, so one just considers 
specific examples.
Within the (of course necessarily limited) knowledge
of the author, this phenomenon has been extensively
discussed in ref.\cite{Bender-Cooper} and in ref.\cite{gaveau}.

The time evolution of a particle
in the symmetric double-well potential
(see fig.~\ref{fig_pmerzbacher})
\beq
\label{V_simm}
V(x) \, = \, \frac{\omega^2}{2} \, 
\big( |x| - a \big)^2; 
\qquad \omega > 0; \quad a \ge 0;	
\eeq
whose spectrum had been calculated in ref.\cite{merzbacher},
was re-considered in ref.\cite{Bender-Cooper}. 
By numerically solving the Schrodinger evolution of this system 
with the particle initially (at $t=0$) put in the left cavity, 
it was found that it does exist a sequence of times at which the particle sits in the right cavity, i.e. the transition 
from the left cavity to the right one is complete.
The symmetric potential in eq.(\ref{V_simm}) was then generalized to various asymmetric potentials with the right cavity generally larger and deeper than the left one.
For some asymmetric potentials, it was found 
that it does not exist any time at which 
the transition to the right cavity is complete.
  
In ref.\cite{gaveau} resonance dynamics at Finite Volume
or, equivalently, in the Quasi-Continuum limit, 
was studied in abstract form by means of a Hamiltonian
operator $\hat{H}$ given by a big Hermitian matrix
of the form
\beq
\label{H_matrix}
H \, = \, \big( H_{i,j} \big)_{i,j=0,1,2,\cdots,N}
\, \equiv \, 
\left(
\begin{array}{ccccc}
\epsilon & g_1 & g_2 & \cdots & g_n 
\\
\bar{g}_1 &  \omega_1 & 0 & \cdots & 0
\\
\bar{g}_2 & 0 & \omega_2 & \cdots & 0
\\
\cdots & \cdots & \cdots & \cdots & \cdots
\\
\bar{g}_N & 0 & 0 & \cdots & \omega_N
\end{array}
\right) ;
\eeq
where the over-lying bar denotes complex conjugation.
The energies on the main diagonal are all real,
\beq
\epsilon, \,\, 
\omega_1, \,\,  \omega_2, \,\, \cdots, \,\,  \omega_N
\, \in \, \RR.
\eeq
The matrix $H$ above, of order $N+1$, describes the coupling of a
discrete level with energy $H_{0,0} = \epsilon$
to the Q.C. states with energies
$H_{i,i} = \omega_i$ by means of the (in general complex) 
couplings given by the off-diagonal elements
$H_{0,i} = g_i$ $(i=1,2,\cdots,N)$.
Without any generality loss, the energies of the Q.C. states 
can be assumed to be ordered,
\beq
\omega_1 \,  \le \, \omega_2 \, 	\le \, \cdots \, \le \, \omega_N.
\eeq
Furthermore, the discrete state energy $\epsilon$ is 
assumed to lie between the Q.C. ones,
\beq
\epsilon \, \in \, \left[ \omega_1, \, \omega_N \right],
\eeq
or to be close to them.
By taking a proper distribution of the couplings $g_i$ 
(sufficiently large and not too smooth at thresholds)
and making proper choices for the diagonal elements 
$\epsilon$ and $\omega_i$,
a limited quantum decay was observed at all times,
together with the general time recursion
phenomenon.
Let us make a few remarks about this model.
\begin{enumerate}
\item
There is not any direct coupling 
of the Q.C. states among themselves%
\footnote{
The $N \times N$ sub-matrix obtained by canceling the first
row and the first column of $H$ in eq.(\ref{H_matrix}) 
is a diagonal matrix.
}.
That implies that, by switching off all the couplings, 
$g_i \to 0$, $i=1,2,\cdots,N$,
each Q.C. state $\omega_i$, 
as well as the discrete state $\epsilon$, 
"evolves diagonally", i.e. it becomes an eigenstate of $H$.
\item
The matrix $H$ given in eq.(\ref{H_matrix}) is actually the 
generalization to the Quasi-Continuum case of the 
Continuum Hamiltonian model originally introduced 
in ref.\cite{fano}.
\end{enumerate}
In the study of resonance dynamics in the Q.C. 
limit, i.e. at finite but large volume,
it would also be interesting to determine
whether there exist some temporal regions
where the decay is, within a reasonable
approximation, exponential in time,
as well as to investigate whether the Zeno effect
still occurs at small times.
For very large volumes, 
where extremely large recursion times are expected,
if an exponential region is found,
one could also try to identify a post-exponential, 
power-decay region, if any.


\section{Standard Winter Model}
\label{stand_wint_mod}

By standard Winter model we mean the usual 
Winter model at infinite volume or,
if we work in momentum space, in the 
Continuum.
The Hamiltonian operator of the model 
and the boundary conditions on the wavefunctions
have been given in eq.(\ref{H_Winter}) and 
in eq.(\ref{BC_Winter}) of the Introduction
respectively, so we do not repeat them here.
As far as dynamics is concerned,
we may say that 
the particle undergoes a perfect, i.e. complete, 
reflection at the space origin, $x=0$, 
and an almost-perfect reflection at the potential barrier, 
$x=\pi$, if the coupling is small, $0 < g \ll 1$
(for simplicity's sake, we restrict ourselves to
a repulsive interaction).
In each collision of the particle with the barrier
at $x=\pi$, in addition to a large reflected wave, 
a small transmitted wave is also generated because 
of tunneling effect%
\footnote{
As well known, if we restrict ourselves to a set of 
eigenfunctions with an upper bound on the energy,
the Dirac $\delta$-potential $\delta(x-\pi)$
of Winter model can be well approximated by a high 
and thin rectangular potential of unit area.
}.


\subsection{Strong-Coupling Limit}

In the strong-coupling limit of the model,
\beq
g \, \to \, + \, \infty, 
\eeq
the potential barrier at $x=\pi$ disappears,
so that
\beq
\hat{H}_{g=\infty} \, = \, - \, \frac{\partial^2}{\partial x^2},
\eeq
and the system reduces to a free particle on the half-line 
\beq
0 \, \le \, x \, < \, \infty.
\eeq
The latter, as well known, possesses a continuous, 
non-degenerate energy spectrum with eigenfunctions
\beq
\Phi_k(x;t) \, = \, 
\sqrt{\frac{2}{\pi}}\sin\left(k x\right) \, e^{-ik^2 t};
\qquad k \, \in \, \RR^+.
\eeq
We have assumed the (conventional) continuum normalization:
\beq
\left\langle \Phi_p \big| \Phi_k \right\rangle
\, = \, \delta(p-k); \qquad p,\, k \, \in \, \RR^+.
\eeq


\subsection{Weak-Coupling Limit}

In the free limit, 
\beq
g \, \to \, 0^+,
\eeq
it turns out that the potential barrier of Winter model
becomes impenetrable%
\footnote{
The free limit $g \to 0^+$ of the potential
barrier $V(x)=\delta(x-\pi)/(\pi g)$ can be intuitively 
thought as a rectangular-shaped potential with infinite area,
such as for example a rectangle with a small, finite thickness (base) 
and infinite height.
}.
Since the Hamiltonian in eq.(\ref{H_Winter}) 
has a pole for $g \to 0$,
this limit has to be studied indirectly,
by calculating its eigenfunctions as
functions of $g$ and then taking the free limit on them
\cite{primo,winter,secondo,lungo}.
In complete agreement with physical intuition,
as anticipated in the Introduction,
for $g \to 0^+$ the system reduces to the union
of two non-interacting subsystems:
\begin{enumerate}
\item
{\bf a free particle in the box $[0,\pi]$,}
having a non-degenerate discrete spectrum
with (normalized to one) eigenfunctions
\beq
\phi_n(x;t) \, = \, 
\sqrt{\frac{2}{\pi}}\sin\left(n x\right) \, e^{-i n^2 t};
\qquad 0 \, \le \, x \, \le \, \pi;
\qquad
n = 1, 2, 3, \cdots.
\eeq
\item
{\bf a free particle in the half line $[\pi,\infty)$,}
possessing a non-degenerate continuous spectrum
with eigenfunctions
\beq
\varphi_k(x;t) \, = \, \sqrt{\frac{2}{\pi}}\sin\big[k (x-\pi)\big] 
\, e^{-i k^2 t};
\qquad
\pi \, \le \, x \, < \, \infty;
\qquad k > 0.
\eeq
The latter are normalized as in the previous section as:
\beq
\left\langle \varphi_p \big| \varphi_k \right\rangle
\, = \, \delta(p-k); \qquad p, \, k \, > \, 0.
\eeq

\end{enumerate}


\subsection{Qualitative Discussion on Resonance Dynamics}

As general unstable state, let us consider 
an initial wavefunction of the particle $(t=0)$ identically
vanishing outside the $[0,\pi]$ resonating cavity:
\beq
\Psi(x; t=0) \, \equiv \, 0
\qquad \mathrm{for} \quad x \, > \, \pi. 
\eeq
One can take for example an eigenfunction of the box $[0,\pi]$
(which is nothing but the $[0,\pi]$ resonating cavity with 
an impenetrable barrier at $x=\pi$ also),
continued to zero outside the box:
\beq
\label{box_eigen}
\Psi^{(h)}(x; t=0) 
\, \equiv \,
\left\{
\begin{array}{cc}
\sqrt{2/\pi} \sin(h x) ; & 0 \, \le \, x \, \le \, \pi ;
\\
           0 ;           & \, \pi \, < \, x \, < \, \infty ;
\end{array}
\right.
\qquad h = 1,2,3,\cdots.
\eeq
Since the above wavefunctions form a basis for
any function $f(x)$ defined for $x \in [0,\infty)$
with support in $[0,\pi]$ only, any initial wavefunction
$\Psi(x; t=0)$ possesses an expansion in terms of the 
set $\left\{ \Psi^{(h)}(x; t=0) ; \, h = 1,2,3, \cdots \right\}$.
Therefore there is not any generality loss in considering
the box eigenfunctions (eq.(\ref{box_eigen})) 
individually, i.e. one at a time. 

The time evolution of such a wavefunction can be qualitatively
described as follows.
Since the potential barrier is slightly penetrable
at small coupling, $0 < g \ll 1$, 
a small component of the initial wavefunction
filters through $x=\pi$ because of tunneling effect,
so that an outgoing wave is generated at $x>\pi$, 
representing the "decay products" of the unstable
state under consideration.
Because time evolution is unitary, the outgoing wave
is generated "at the expense" of the inside amplitude,
i.e. the inside amplitude is diminished.
Since the outgoing wave does not hit any barrier
in its propagation to the right, far from the cavity,
towards $x = + \infty$,
the inside amplitude is never regenerated by the outgoing
wave, so that the no-decay probability,
\beq
P_{ND}^{(h)}(t) 
\, \equiv \, 
\left|
\int\limits_0^\pi \bar{\Psi}^{(h)}(x;0) \,  e^{-i\hat{H} t } \, \Psi^{(h)}(x;0) \, dx
\right|^2;
\qquad
h = 1, 2, 3, \cdots;
\eeq
is basically a monotonically-decreasing function of time
down to zero, in agreement with the general theory.


\section{Winter Model at Finite Volume}
\label{wint_mod_fin_vol}

We generalize the standard Winter model 
considered in the previous section,
by restricting the particle coordinate $x$
to a segment,
\beq
0 \, \le \, x \, \le \, L \, < \, \infty, 
\eeq
and assuming a vanishing (or reflecting) boundary 
condition also at the final endpoint of the segment 
(see fig.~\ref{fig_pWinterfi}):
\beq
\psi(x = L; \, t) \, \equiv \,0;
\qquad t \, \in \, \RR.
\eeq
The model so constructed describes a small resonating cavity 
($x \in [0,\pi]$)
weakly-coupled for $|g|\ll 1$ to a large one ($x \in [\pi,L]$);
the resonant states of the small cavity are coupled to close, 
equally-spaced momentum levels.
Let us remark that, while standard Winter model is a 
one-parameter model, its extension at finite volume
is a two-parameter model --- namely $g$ and $N$.
However, even after the extension to finite volume,
the model remains relatively simple, so that computations 
can still be made to some extent 
--- as we are going to show --- analytically.
Winter model at infinite volume is recovered 
from the finite-volume model by taking the limit 
$L \to + \infty$.


\subsection{Qualitative Discussion on Resonance Dynamics}

Similarly to the infinite-volume case,
let us consider an initial wavefunction
identically vanishing outside the small
cavity $[0,\pi]$, i.e. exactly zero inside 
the large cavity, for $x \in [\pi,L]$.
As in the infinite-volume case, for $0 < g \ll 1$, 
a little outgoing wave is generated at 
small evolution times out of the initial amplitude.
By increasing time from zero up to
\beq
t \, \approx \, t_{one-way} \, \equiv \, \frac{L}{v}
\qquad
(L \gg \pi);
\eeq
where $v$ is the average group velocity of the wavepacket,
the outgoing wave propagates up to the right border of the
large cavity, at $x=L$, where it is (completely) reflected
back in the allowed region $[0,L]$.
Such a wave, by propagating to the left,
goes back towards the small cavity and then
suffers at the point $x=\pi$ a large reflection 
and a small transmission, at the time 
\beq
t \, \approx \, t_{two-way} \, \equiv 2 \,t_{one-way} 
\, \equiv \, \frac{2 L}{v}.
\eeq
As in any collision of the particle with the
barrier at $x=\pi$, there is a small
transmitted wave because of the small 
permeability of the latter. 
A small wave component therefore goes back 
inside the small cavity.
Such a "regeneration phenomena"
--- which we have described pictorially in the Introduction
with the $Z^0$ example --- is characteristic of resonances at 
finite volume, as it clearly does not occur at infinite volume.
The general phenomenon of time recurrence 
(which we have discussed in sec.$\,$\ref{sec_recur}) 
can be explained qualitatively for the present model
in terms of the above dynamical mechanism:
by waiting a proper, sufficiently long,
evolution time, multiple reflection gives rise to
an almost complete re-entering of the
initial wavefunction inside the small cavity.
Also the phenomenon of limited decay 
(which we have discussed in sec.$\,$\ref{lim_decay}),
if it actually occurs 
in some parameter-space region
of the finite-volume Winter model, 
can be explained qualitatively as follows.
Let us expand the initial wavefunction
of the unstable state in eigenfunctions 
of $\hat{H}$, the complete, interacting 
Hamiltonian of the system.
Since the different energy components
have different transmission and reflection
coefficients (the latter depend upon the energy), 
as well as different propagation velocities, 
it may happen that for no times all these components
are outside of the small cavity,
so that the no-decay probability never
vanishes.


\section{Spectrum}
\label{spec_vol_fin}

Unlike standard Winter model,
which has a continuous spectrum,
finite-volume Winter model
has a discrete spectrum which,
as discussed in the Introduction, 
is naturally decomposed in an exceptional part,
which we consider first, and an ordinary part.


\subsection{Exceptional Eigenvalues and Eigenfunctions}

\begin{figure}[ht]
\begin{center}
\includegraphics[width=0.5\textwidth]{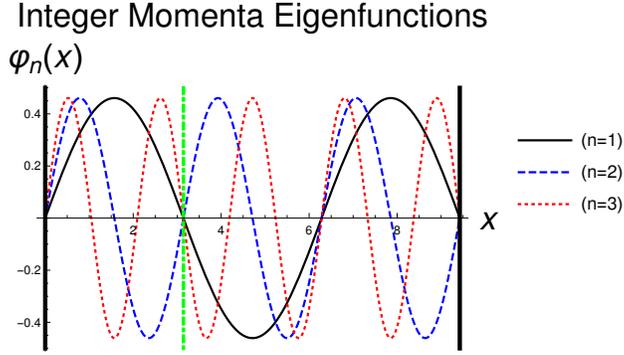}
\footnotesize
\caption{
\label{fig_pefecc}
\it Plots of the first three exceptional eigenfunctions of 
finite-volume Winter model with the exactly integer
momenta $p=1,2,3$ for $L=3\pi$, i.e. for
the large cavity (to the right of the vertical thick green
dot-dashed line) two times bigger 
than the small one (to the left of the green line).
The boundaries of the system are represented
by black thick continuous vertical lines.
}
\end{center}
\end{figure}

It is immediate to check that the wavefunctions
\beq
\label{eigen_k_integer}
\varphi_n(x) \, = \,
\sqrt{\frac{2}{L}} \, \sin\left(p_n \, x\right);
\qquad
0 \, \le \, x \, \le \, L;
\eeq
having the exactly-integer momenta,
\beq
p_n \, \equiv \, n \, = \, 1, 2, 3, \cdots,
\eeq
are eigenfunctions of the finite-volume Winter model
(see fig.~\ref{fig_pefecc}).
The quantity $L$, the total length of the system,
is assumed, for simplicity's sake, to be an
integer multiple of $\pi$, i.e. of the length
of the small cavity,
\beq
L \, = \, M \pi; \qquad
M \,\ \mathrm{integer}.
\eeq
%
%
%
Since all the $\varphi_n(x)$'s exactly vanish at $x=\pi$,
\beq
\varphi_n(x=\pi) \,  \equiv \, 0;
\qquad n \, \in \, \NN ;
\eeq
the $\delta$-potential, with support at this point only,
does not have any effect on the matching condition
on the eigenfuctions.
The consequences of this fact are the following:
\begin{enumerate}
\item
The exceptional eigenvalues $p_n$ and 
the exceptional eigenfunctions $\varphi_n$
do not have any dependence on the coupling of 
the model $g$.
\item
The eigenfunctions $\varphi_n(x)$ in eq.(\ref{eigen_k_integer})
are smooth functions of $x$, i.e. they
have continuous space derivatives of all
orders in the interval $(0,L)$,
\beq
\varphi_n \,  \in \, C^\infty(0,L);
\qquad n \, \in \, \NN ;
\eeq
\item
If the $\varphi_n$'s are restricted to the interval 
$[0,\pi]$, they become eigenfunctions of the free particle
in the box $[0,\pi]$ (actually, all the box eigenfunctions
are obtained this way).
Furthermore, if the $\varphi_n$'s are restricted to the 
large cavity $[\pi,L]$,
they become eigenfunctions of the free particle
in the box $[\pi,L]$.
\end{enumerate}
Let us end this section by noting that
if the $\varphi_n(x)$'s are extended smoothly to the
entire positive half-line,  $x \in [0,\infty)$,
they become eigenfunctions of the standard Winter model.
However, since the spectrum of the usual model is continuous,
these eigenfunctions form a negligible, zero-set measure.


\subsection{Normal Eigenvalues}

The quantum number $k>0$, appearing in the
normal eigenfunctions $\psi(k;x)$ (see later), is a 
quantized, non (exactly) integer momentum,
satisfying the real transcendental equation
\beq
\label{eq_basic}
\cot(\pi k) \, + \, \cot(\pi N k) 
\, + \, \frac{1}{\pi g k} \, = \, 0
\qquad (g \ne 0);
\eeq
where the integer
\beq
N \, \equiv \, M - 1
\eeq
is the length, in units of $\pi$
of the large cavity 
\beq
[\pi,\, L] \, = \, [\pi, \, M \pi] 
\, = \, \left[\pi, \, (N+1) \pi \right].
\eeq
Since all the terms appearing in
eq.(\ref{eq_basic}) are odd functions of $k$, it follows
that, if $\bar{k}$ is a solution of eq.(\ref{eq_basic}), 
then also $-\,\bar{k}$ is a solution of this equation.
This fact is a consequence of the fact that the 
vanishing boundary condition at $x=0$, 
eq.(\ref{BC_Winter}), 
is a reflecting one, so that momentum is conserved only 
up to a sign.
Without any generality loss, we can therefore assume
\beq
k \, > \, 0.
\eeq
We may say that solving eq.(\ref{eq_basic})
--- determining all the normal component of the momentum
spectrum of finite-volume Winter model ---
is the main aim of this paper.
We have not been able to solve this equation analytically 
in terms of known special functions, so we will resort to 
numerical and perturbative methods.
An important point is to label the allowed momenta $k$
in a physically appealing way.
Since for $g \to 0$, i.e. in the free limit of the model, 
the last term on the l.h.s. of
eq.(\ref{eq_basic}) diverges, it follows
that the sum of the cotangent functions
have to diverge also.
Since $N$ is integer, the poles of $\cot(\pi k)$
are also poles of $\cot(\pi N k)$.
That implies that also the function $\cot(\pi N k)$ 
diverges in the free case, so that 
the allowed momenta $k$ must have limiting values
\beq
k^{(0)} \,  = \, \frac{s}{N};
\qquad
s \, = \, 1, 2, 3, \cdots
\qquad
(g \to 0). 
\eeq
By the general principle of adiabatic continuity \cite{anderson},
let us therefore label all the normal momenta $k$,
in the general interacting case $g \ne 0$, by the value
taken for $g \to 0$:
\beq
k \, = \, k_{s/N}(g)
\eeq
with
\beq
\lim_{g \to 0} k_{s/N}(g) \, = \, \frac{s}{N};
\qquad
s \, = \, 1, 2, 3, \cdots;
\quad k \, \in \, \RR^+.
\eeq


\subsubsection{Strong-Coupling Limit}

For $g \to + \, \infty$, i.e. in the strong-coupling limit, 
the momentum spectrum equation (\ref{eq_basic}) basically simplifies 
into the elementary equation
\beq
\sin(\pi M k) \, = \, 0;
\eeq 
where we remember that
\beq
M \, = \, N + 1
\eeq
is the total length of the system
(small cavity $+$ large cavity) 
in units of $\pi$.
The solutions of the above equation are
the momenta (see figs.$\,$\ref{fig_pNsmall}$-$\ref{fig_pNlarge})
\beq
\label{k_strong_coupl}
k \, = \, \frac{s}{M},
\qquad s = 1,2,3,\cdots.
\eeq
For $g \to \infty$, the potential barrier,
given by the $\delta$-function,
completely disappears and the system reduces
to a free particle in the box $[0,L]$.
The level spacing in this limit 
approaches the constant value
\beq
\label{strong_coupl_spac}
\Delta k_{g=\infty} \, = \, \frac{1}{N+1}.
\eeq
Therefore, in the strong-coupling limit,
the momentum spectrum
of finite-volume Winter model goes into
a non-degenerate, equally-spaced spectrum.
When the index $s$ is an integer multiple of $M$, then 
the momentum in eq.(\ref{k_strong_coupl}) is integer:
\beq
s \,= \, n \, M \quad \Rightarrow \quad k \, = \, n;
\qquad n = 1, 2, 3, \cdots;
\eeq
and we re-obtain the exceptional momenta.


\subsubsection{Free-Theory Limit}
\label{sec_free_lim}

\begin{figure}[ht]
\begin{center}
\includegraphics[width=0.5\textwidth]{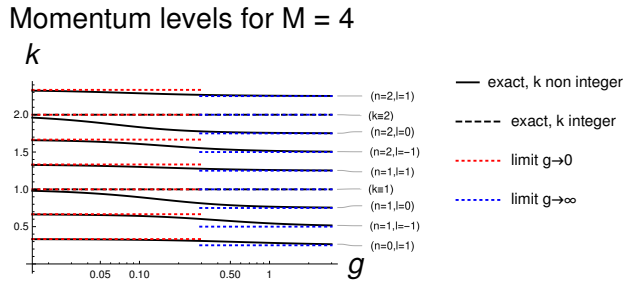}
\footnotesize
\caption{
\label{fig_pNsmall}
\it plots of the first nine momentum levels  
$k$ as functions of the coupling $g$, 
$k = k(g)$ (the continuous and dashed black lines), 
of Winter model at $M=4$,
i.e. for the system with the large cavity three times 
bigger than the small one. 
The red dotted lines represent the free limit
of the corresponding levels $\left(\lim_{g \to 0} k(g)\right)$, 
while the blue dotted 
lines represent the strong-coupling limit
$\left(\lim_{g \to \infty} k(g)\right)$.
The degeneracy for $g \to 0$ of the flat level 
$k \equiv 1$ (the lower black dashed line) 
with the resonating level $k_1(g)$ right below it
--- in an otherwise equally-spaced spectrum ---
is clearly visible;
a similar degeneracy occurs between the level $k \equiv 2$
(the upper black dashed line)
and the level $k_2(g)$ right below it.
The level spacing in the free limit, 
$\Delta k_{g=0} = 1/(M-1) \simeq 0.33$,
is given by the distance between two adjacent red lines,
while the spacing in the strong-coupling limit,
$\Delta k_{g=\infty} = 1/M = 0.25$,
is given by the distance between two
adjacent blue lines.
The fact that the free-theory spacing is larger than
the strong-coupling one is clearly visible.
Note that the horizontal scale (that of the coupling $g$) 
is logarithmic.
}
\end{center}
\end{figure}

As discussed above, for $g \to 0^+$, i.e. in the free limit,
the allowed momenta are of the form
(see figs.$\,$\ref{fig_pNsmall}$-$\ref{fig_pNlarge})
\beq
\label{order_zero}
k^{(0)} \, = \, \frac{s}{N}.
\eeq 
By means of the euclidean division of $s$ by $N$,
\beq
\label{div_eucl}
s \, = \, n \, N \, + \, l,
\eeq
the free-theory momentum can also be written
in the form
\beq
\label{k_0_nonres}
k^{(0)} \, = \, n \, + \, \frac{l}{N}.
\eeq
We take the remainder $l$ in the 
(quasi-)symmetric range
\beq
- \, \frac{N}{2} \, < \, l \, \le \, \frac{N}{2}.
\eeq
In the following, a momentum level $k$, considered
as a function of the coupling $g$: $k=k(g)$,
which is equal to $n+l/N$ in the limit $g \to 0$,
will be called a $(n,l)$ level or more simply
a $l$ level.
When $s$ is an integer multiple of $N$, then the
momentum in eq.(\ref{order_zero}) is integer:
\beq
\label{k_ex_integer}
s \, = \, n \, N \quad \Rightarrow \quad k \, = \, n;
\qquad n = 1, 2, 3, \cdots;
\eeq
and we re-obtain the exceptional momenta.
The level spacing in the free limit
is given by:
\beq
\Delta k_{g=0} \, = \, \frac{1}{N}.
\eeq
Note that it is larger than the strong-coupling spacing 
(cfr. eq.(\ref{strong_coupl_spac})). 
As we are going to show in the next section,
the levels $k_n(g)$ become degenerate 
with the exceptional levels $p_n \equiv n$
in the limit $g \to 0$ 
and this degeneracy "compensates" for the larger 
$g\to 0$ spacing, giving rise to the same average 
level density as for $g \to \infty$. 

To summarize, in the free limit, the momentum spectrum
of finite-volume Winter model goes into
an equally-spaced spectrum, eq.(\ref{order_zero}), 
with a double degeneracy at integer momenta
(eq.(\ref{k_ex_integer})).

%
%


\subsubsection{Discussion}

\begin{figure}[ht]
\begin{center}
\includegraphics[width=0.5\textwidth]{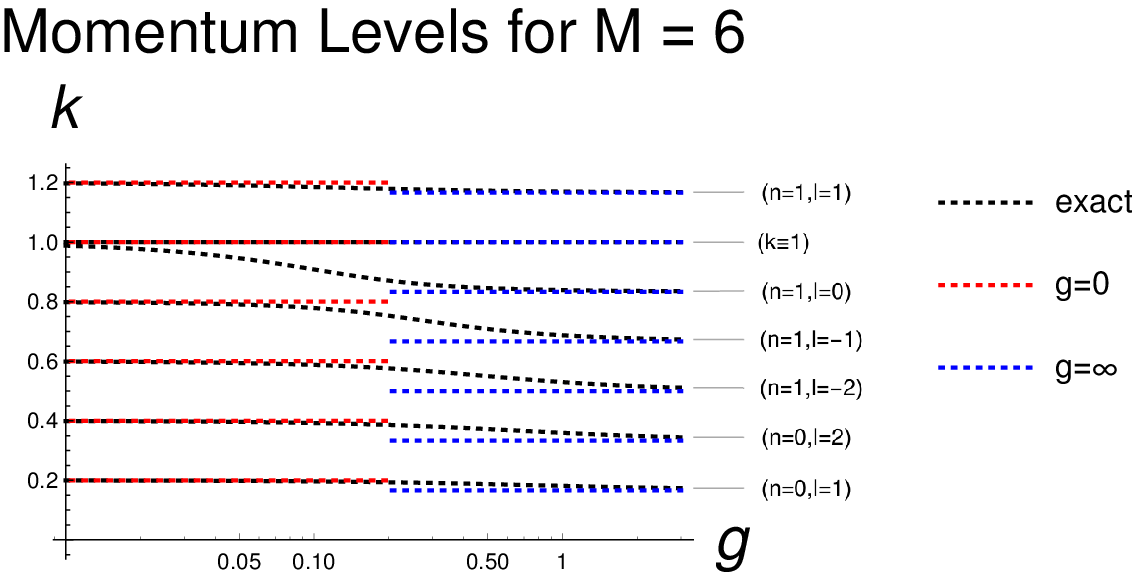}
\footnotesize
\caption{
\label{fig_pNsmall2}
\it lowest seven momentum levels of Winter model,
containing the first resonance, for $M=6$.
The black dotted lines represent the exact levels,
the red dotted lines the free limit ($g \to 0$)
and the blue dotted lines the strong-coupling limit 
($g \to \infty$).
The degeneracy for $g \to 0$ of the flat level 
$k \equiv 1$ with the level just below it, namely
the resonant level $k_1(g)$, is clearly visible.
One may also notice that the level spacing at $g=0$, 
given by $\Delta k_{g=0} = 1/(M-1) = 0.2$,
is larger than the spacing at $g=\infty$, 
which is equal to $\Delta k_{g=\infty} = 1/M \simeq 0.167$.
As in previous plot, the horizontal scale is logarithmic.
}
\end{center}
\end{figure}

In figs.$\,$\ref{fig_pNsmall} and \ref{fig_pNsmall2}
we plot the lowest momentum levels $k$ as functions 
of the coupling $g$, $k = k(g)$,
of finite-volume Winter model in the small-volume cases 
$M=4$ and $M=6$ respectively.
The curves shown are obtained through numerical solution 
of eq.(\ref{eq_basic}).
%
%
After excluding the flat levels, 
corresponding to the exceptional eigenfunctions 
in eq.(\ref{eigen_k_integer}),
we observe that each momentum level $k$ is a 
strictly monotonically-decreasing 
function of $g$, with an upper horizontal
asymptote given by the free limit $(g \to 0)$,
represented by a dotted red line,
and a lower horizontal asymptote given by the
strong-coupling limit $(g \to \infty)$,
represented by a dotted blue line.

By approaching a flat level from below,
one encounters levels with total variation
(the distance between the red
line and the lower blue line) 
progressively bigger;
%
\begin{figure}[ht]
\begin{center}
\includegraphics[width=0.5\textwidth]{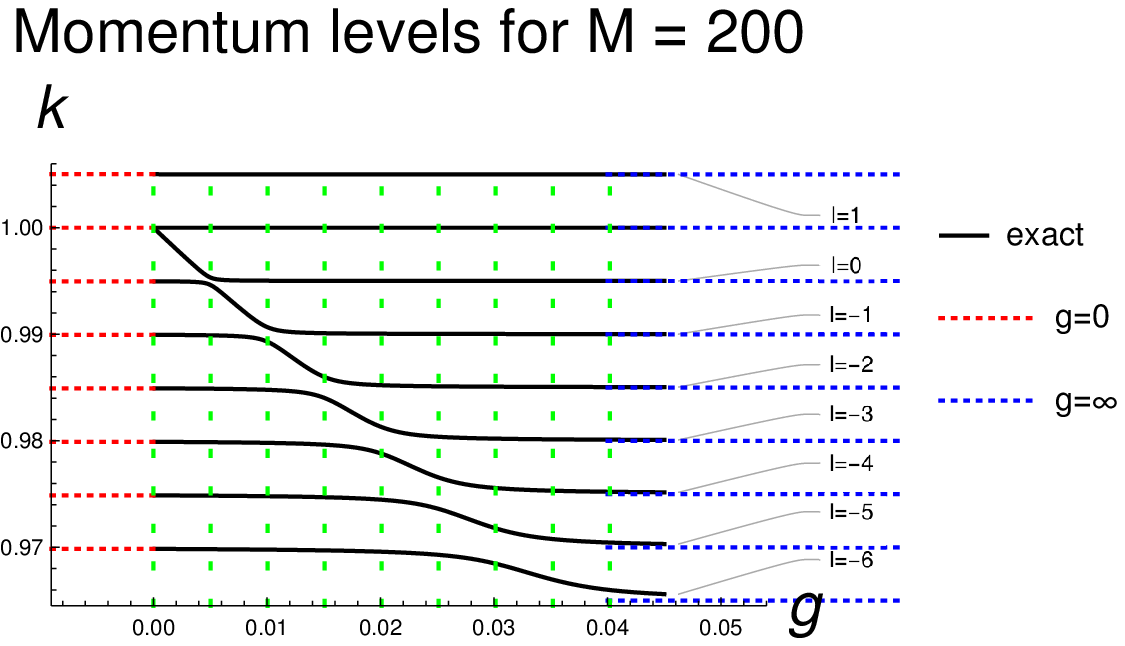}
\footnotesize
\caption{
\label{fig_pNlarge}
\it Momentum levels around the first or
fundamental resonance $(n=1)$ for volume $M=200$.
The black continuous lines represent the exact levels,
the red dotted lines the free limit $g \to 0$
and the blue dotted lines the strong-coupling limit 
$g \to \infty$.
The vertical light green dashed lines at 
$g = j/199 \cong 0.005 \times j$ $(j=0,1,2,3,\cdots)$
represent the coupling values where
exactly-degenerate doublets $(j=0)$ or 
quasi-degenerate doublets $(j>0)$ occur in the spectrum.
}
\end{center}
\end{figure}
%
the total variation of the level 
right above a flat one is instead 
extremely small.
As already discussed, a flat level separates the
levels having a resonant behavior (below it)
from the levels not having such a behavior
(above it).
%
\begin{figure}[ht]
\begin{center}
\includegraphics[width=0.5\textwidth]{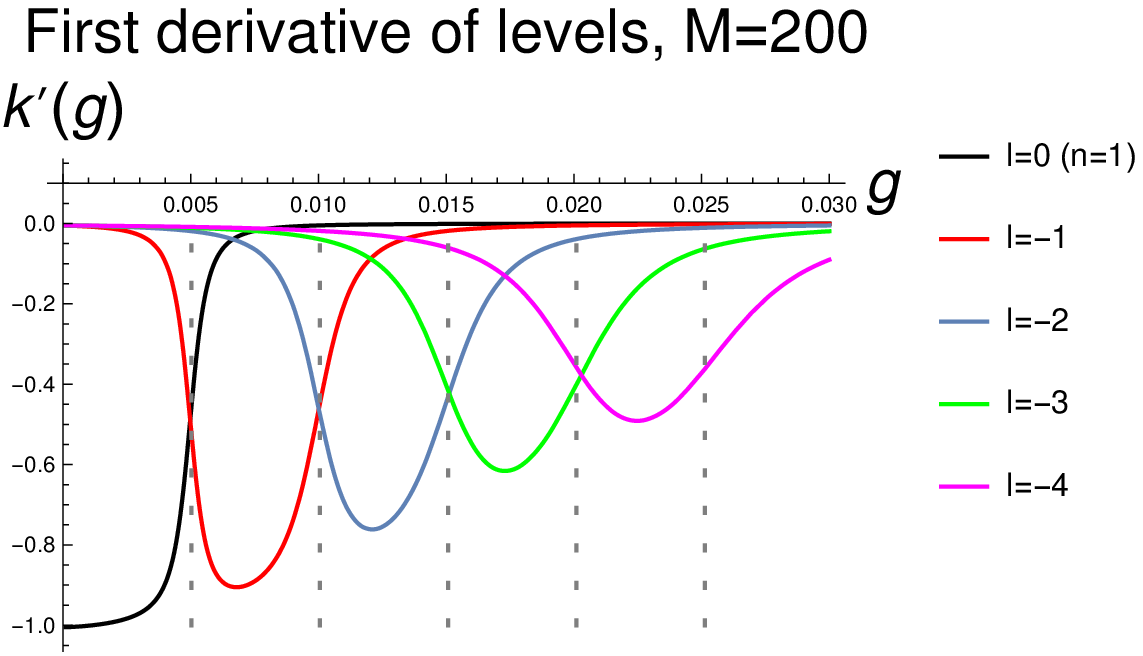}
\footnotesize
\caption{
\label{fig_pfirstder}
\it first derivatives of momentum levels $k$
as functions of the coupling $g$, namely $k'(g)$, 
below and including the first or
fundamental resonance $(n=1)$ for $M=200$.
The vertical gray dashed lines at 
$g = j/199 \cong 0.005 \times j$ $(j=1,2,3,\cdots)$
represent the coupling values where 
quasi-degenerate doublets occur in the spectrum.
Note that the intervals around the minima of $k'(g)$
lie between two consecutive dashed lines.
}
\end{center}
\end{figure}
%
%
A degeneracy for $g \to 0$ of the level 
$k \equiv 1$ with the level right below it
(whose physical origin has been explained in 
sec.$\,$\ref{sec_free_lim})
is clearly seen in both figs.$\,$\ref{fig_pNsmall} and \ref{fig_pNsmall2}; 
A similar degeneracy of the level $k \equiv 1$
with the level right below it 
is also visible in fig.~\ref{fig_pNsmall2}. 
%
\begin{figure}[ht]
\begin{center}
\includegraphics[width=0.5\textwidth]{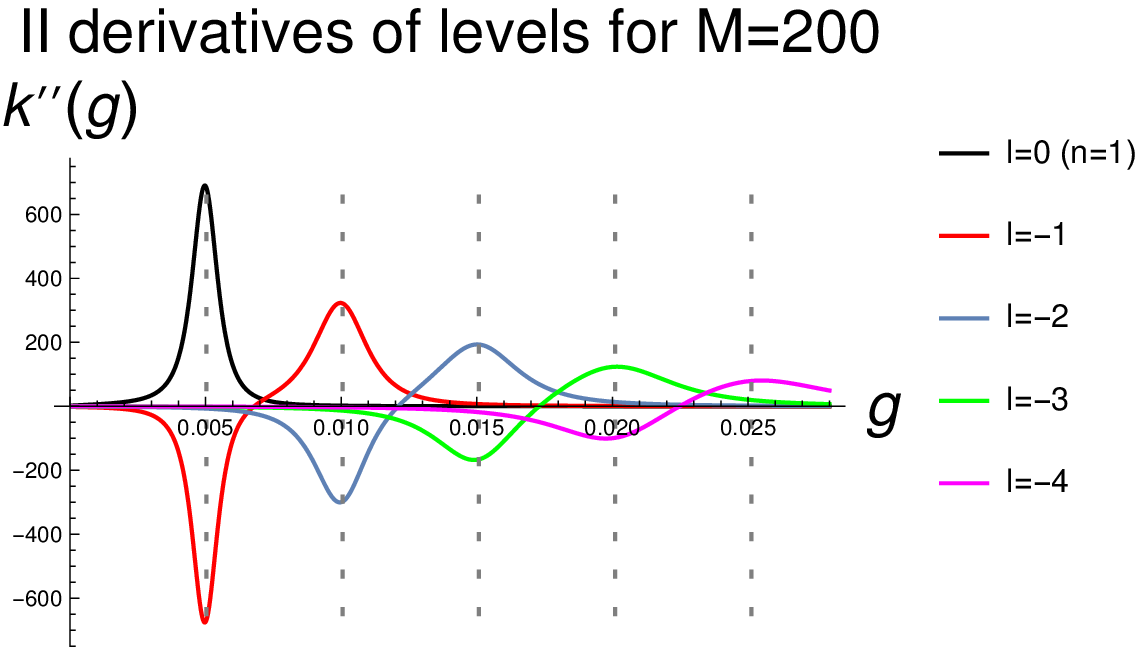}
\footnotesize
\caption{
\label{fig_psecder}
\it second derivatives of momentum levels $k$
as functions of the coupling $g$,
$k''(g)$, below and including the first or
fundamental resonance $(n=1)$ for $M=200$.
The maxima and the minima of the function $k''(g)$
are quite close to the gray vertical dashed lines.
}
\end{center}
\end{figure}

%
\begin{figure}[ht]
\begin{center}
\includegraphics[width=0.5\textwidth]{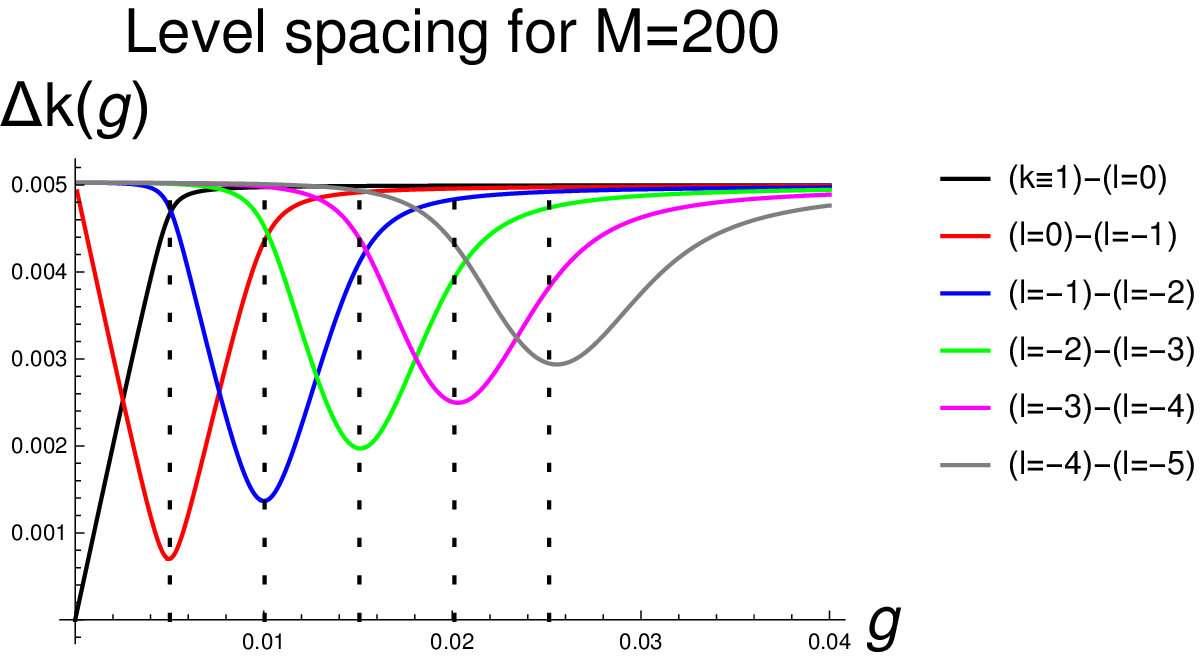}
\footnotesize
\caption{
\label{fig_pdifferenza}
\it level spacing $\Delta k$
(difference between two contiguous momenta) 
as function of the coupling $g$,
$\Delta k = \Delta k(g)$, for the
levels below and including the first or fundamental resonance 
$(n=1)$ for $M = 200$.
The minima of $\Delta k(g)$
occur at $g \approx g_j \cong 0.005 \times j$, $j=0,1,2,3,\cdots$,
pretty close to the vertical gray dashed lines.
By increasing $j$, the minima get less pronounced.
}
\end{center}
\end{figure}
%


In fig.~\ref{fig_pNlarge} we plot the momentum
levels around the first or fundamental resonance 
$(n=1)$ in the large-volume case $M=200$.
We notice that the resonant level $l=0$
(the level right below the flat one $k \equiv 1$)
goes down roughly in a linear way with the coupling 
from the free-theory point $g=0$
(where it is degenerate with $k \equiv 1$,
as we have already seen in the small-$M$ cases above)
up to 
\beq
g \, \approx \, g_1 \, \equiv \, \frac{1}{N} \, \cong \, 0.005
\qquad (N=199); 
\eeq
where it shows a sharp variation of its first derivative ---
a quick increase up to zero; 
this phenomenon is also clearly seen both in figs.$\,$\ref{fig_pfirstder} and \ref{fig_psecder}.
For $g \gsim g_1$, this level $(l=0)$ is roughly constant,
having already reached basically its infinite-coupling limit. 

In general, the levels with $l<0$ (below the resonant one)
are pretty flat for very small and 
very large couplings and show a rather sharp transition at 
$g = \mathcal{O}(1/N)$, when they substantially go from 
their respective free-theory values down to their 
infinite-coupling values.
The $l=-1$ level, for example, 
is basically constant from $g=0$ up to
$g \approx g_1$, where it shows
a sharp decrease of its first derivative.
In the coupling interval
\beq
g_1 \,\, \lsim \,\, g \,\, \lsim \,\, 
g_2 \, \equiv \, \frac{2}{N} \, \cong \, 0.01,
\eeq
this level goes down linearly with $g$.
At $g \approx g_2$, it shows a sharp increase of its 
first derivative up to zero
and finally, for $g \gsim g_2$, it is
roughly constant, having basically reached 
its infinite-coupling limit.
In general, the non-resonant level of index $l=-1,-2,-3,\cdots$
is constant from $g=0$ up to roughly 
\beq
g_{|l|} \, \equiv \, \frac{|l|}{N} \, \cong \, 0.005 \times |l|, 
\eeq
where it shows a sharp decrease of its first derivative.
Then, in the coupling interval
\beq
g_{|l|} \,\, \lsim \,\, g \,\, \lsim \,\, 
g_{|l|+1} ,
\eeq
the level goes down roughly linearly with $g$.
For $g \approx g_{|l|+1}$, it shows a sharp
increase of its first derivative up to zero
and for larger couplings it is basically flat.
In general, by going away from a resonance level $(l=0)$ 
by taking $l = -1,-2,-3,-4,\cdots$, 
the above resonance behavior is softened.

Since the transition regions of different $l$
levels are shifted with respect to each other, 
for values of the coupling 
\beq
\label{coup_eccez}
g \, \approx \, g_j \, \equiv \, 
\frac{j}{N} \, \cong \, 5 \times 10^{-3} \times j;
\qquad
j = 1, 2, 3, \cdots;
\eeq
two black curves almost get in touch in fig.~\ref{fig_pNlarge}, 
implying that the corresponding doublet
is quasi degenerate. 
The coupling values $g_j$
correspond to the vertical green dashed lines in 
figs.$\,$\ref{fig_pNlarge}$-$\ref{fig_pdifferenza}.

For the exceptional values of the coupling $g = g_j$ 
(eq.(\ref{coup_eccez})),
the first resonance of the infinite-volume
model corresponds to the quasi-degenerate doublet
having momenta, in the free limit $g \to 0$,
given by $1-(j-1)/N$ and $1-j/N$.
Note that the distance between the quasi-degenerate levels 
rapidly grows with $j$, so that this approximate
degeneracy is barely visible for,
let's say, $j = \mathcal{O}(5)$.

Between the first two vertical green dashed lines
in fig.~\ref{fig_pNlarge}, 
i.e. in the coupling range
\beq
0 \,\, \lsim \,\, g \,\, \lsim \,\, g_1,
\eeq
there are {\it three} levels
--- rather than {\it two} as in general ---
in the momentum interval
\beq
0.995 \,\, \lsim \,\, k \,\, \lsim \,\, 1,
\eeq
of width
\beq
\Delta k \, \simeq \, 5 \times 10^{-3}  
\, \simeq \, \frac{1}{N} \, \simeq \, \frac{1}{N+1}.
\eeq
So long as $j=1,2,3,\cdots$ is not too large, 
this "compression" of three levels
also occurs in the coupling intervals
\beq
g_j 
\,\,\, \lsim \,\,\, g \,\,\, \lsim \,\,\,
g_{j+1} .
\eeq
For large $j$, i.e. for large values of the coupling $g$
compared to $1/N$, a resonance corresponds to a mild
compression of many momentum lines.

We may summarize the above findings by saying that,
for the exceptional values of the coupling 
$g = g_j$ given by eq.(\ref{coup_eccez}), 
the fundamental resonance of the usual
Winter model corresponds, in the finite-volume case, 
to a quasi-degenerate doublet.
For generic values of the coupling,
the resonance corresponds instead to 
a compression of three lines in the
typical spacing between two of them
\beq
\Delta k \, \approx \, \frac{1}{N} \, \approx \, \frac{1}{N+1};  
\eeq
For larger values of the coupling, i.e. for large $j$, 
the resonance corresponds to a mild compression of many levels.
As in the small-volume cases considered above,
only the levels below the flat one $k \equiv 1$
exhibit a resonant behavior; the levels
above the latter 
(we have plotted just one of them in fig.~\ref{fig_pNlarge}) 
do not "see" the resonance and are 
approximately constant in $g$.


\subsection{Normal Eigenfunctions}

\begin{figure}[ht]
\begin{center}
\includegraphics[width=0.5\textwidth]{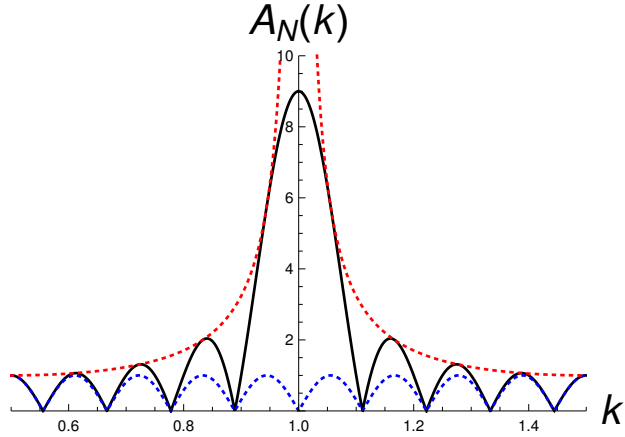}
\footnotesize
\caption{
\label{fig_pinsamp}
\it 
Inside Amplitude $A_N$ as function of momentum $k$,
$A_N = A_N(k) = 
\left|\sin(\pi N k) \right| / \left| \sin(\pi k) \right|$, 
for a complete period, $0.5 \le k \le 1.5$
(black continuous line) for $M=N+1=10$ . 
The red dotted line is the plot of the function
$1/\left|\sin(\pi k)\right|$, namely the denominator of the Inside
Amplitude, while the blue dotted line
is the plot of the function $\left|\sin(\pi N k)\right|$,
the numerator of the Inside Amplitude.
}
\end{center}
\end{figure}

\begin{figure}[ht]
\begin{center}
\includegraphics[width=0.5\textwidth]{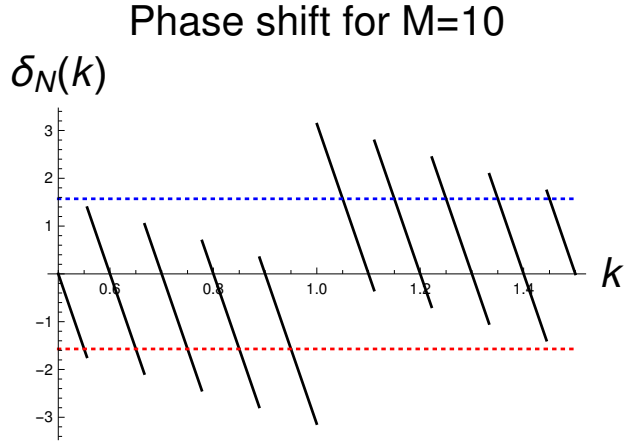}
\footnotesize
\caption{
\label{fig_pfase}
\it Phase Shift $\delta_N$ as function of the momentum $k$,
$\delta_N  = \delta_N(k)$ (black continuous line), 
in the period $0.5 \le k \le 1.5$, for $M=10$.
The blue/red dotted horizontal lines
are the constant phases $\pm \pi/2$
respectively. 
The $2\pi$-discontinuity at $k=1$,
together with the $\pi$-discontinuities
at $k=1+l/N \simeq 1 + 0.11\times l$ 
$(l \ne 0)$, are clearly visible (see text).
}
\end{center}
\end{figure}

As already discussed, by squeezing the system, 
the continuous spectrum of usual Winter model turns into 
a discrete one, with normal eigenfunctions,
as functions of the momentum $k$ (solution of eq.(\ref{eq_basic})),
given by:
\beq
\label{old_ef}
\psi(k; \, x) \, = \,
\mathcal{N}_N(k) \Big\{
\theta(\pi-x) \sin(\pi N k) \sin(kx)
\, + \,
\theta(x-\pi) \sin(\pi k) \sin\big[ k (L - x ) \big]
\Big\} ;
\eeq
where $\theta(z) \equiv 1$ for $z>0$ and zero otherwise
is the Heaviside unit step function. 
As already defined, $N$ is the length, divided by $\pi$,
of the segment $[\pi, L]$,
i.e. the length of the large cavity $[\pi,L]$
to which the small cavity $[0,\pi]$ is coupled 
($g \ne 0$).
The normalization constant has the explicit expression: 
\beq
\mathcal{N}_N(k) 
\, \equiv \, 
\Bigg\{
\sin^2(\pi N k)
\left[
\frac{\pi}{2}
- \frac{\sin(2\pi k)}{4k}
\right]
\, + \,
\sin^2(\pi k)
\left[
\frac{\pi N }{2}
- \frac{\sin(2\pi N k)}{4k}
\right]
\Bigg\}^{-1/2}.
\eeq
In order to investigate resonant effects,
it is convenient to rewrite the above eigenfunctions,
as in the infinite-volume case \cite{winter}, 
in terms of an amplitude and a phase:
\beq
\label{new_ef}
\psi(k; x) \, = \, {\mathcal{C}_N(k)} 
\Big\{
A_N(k) \, \theta(\pi-x) \, \sin(k x)
\, + \,
\theta(x-\pi) \, \sin\big[ k x \, + \, \delta_N(k) \big]
\Big\};
\eeq
where:
\begin{enumerate}
\item
$A_N(k)$ is the ratio  of the inside amplitude 
$(0 \le x \le \pi)$ over 
the outside one $(\pi \le x \le L)$,
\beq
A_N(k) \, = \, \left| D_N(k) \right| ;
\eeq
where (see fig.~\ref{fig_pinsamp}):
\beq
\label{Dir_ker}
\qquad
D_N(k) \, \equiv \, \frac{\sin(\pi N k)}{\sin(\pi k)};
\qquad\qquad
k \, \notin \, \ZZ.
\eeq
Note that, for odd $N$ ($M=N+1$ even), 
the function $D_N(k)$ is the Dirichlet's 
Kernel of Fourier series theory.
\item
$\delta_N(k)$ is the relative phase of the outside amplitude with respect to the inside one 
(see fig.~\ref{fig_pfase}),
\beq
\label{delta_N_def}
\delta_N(k) \, \equiv \, 
\arctan
\left\{
- \, \frac{\sin(\pi k) \cos[ \pi (N+1) k ]}{\sin(\pi N k)} ;
\,\, \frac{\sin(\pi k) \sin\left[\pi (N+1) k\right]}{\sin(\pi N k)}
\right\}.
\eeq
The two-argument arctangent function $\arctan(x,y)$ is a specification
of $\arctan(y/x)$ which also takes into account in which
quadrant the point $(x,y)$ lies in.
\item
${\mathcal{C}_N(k)}$ is an overall normalization constant,
\beq
{\mathcal{C}_N(k)} 
\, \equiv \, 
\Bigg\{
D_N(k)^2
\,
\left[
\frac{\pi}{2}
- \frac{\sin(2\pi k)}{4k}
\right]
\, + \,
\left[
\frac{\pi N }{2}
- \frac{\sin(2\pi N k)}{4k}
\right]
\Bigg\}^{-1/2} .
\eeq
\end{enumerate}
The following remarks are in order:
\begin{enumerate}
\item
Eq.(\ref{new_ef}) differs from eq.(\ref{old_ef})
by an overall sign only, namely 
\beq
\mathrm{sign}\left[D_N(k)\right],
\eeq
which, as well known, is physically irrelevant.
\item
The appearance of the Dirichlet's Kernel $D_N(k)$ 
in the finite-volume eigenfunctions (cfr. eq.(\ref{new_ef}))
suggests that the convergence of the latter
to the corresponding infinite volume ones in the 
limit $N \to \infty$, will have to be intended 
as a weak one.
This fact is a consequence of the rather "hard"
boundary conditions which we
have chosen, namely reflecting ones, in comparison 
to the usual softer periodic boundary conditions.
\end{enumerate}
%
%
Let us then discuss in the following sections 
the main properties of the Inside
Amplitude and of the Phase Shift, as functions
of the (still unrestricted) momentum $k$ .


\subsubsection{Inside Amplitude as function of momentum}

The Inside Amplitude $A_N$ as a function of the
momentum $k$, $A_N = A_N(k)$,
has the following properties:
\begin{enumerate}
\item
{\it Unitary period.}

The Inside Amplitude is periodic
of period $T=1$,
\beq
A_N(k+1) \, = \, A_N(k);
\qquad k \, \in \, \RR.
\eeq
The above equation trivially follows
by taking the modulus on both sides
of the equation (which is immediately verified)
\beq
D_N(k+1) \, = \, (-1)^{N+1} \, D_N(k).
\eeq
Note that the function $D_N(k)$ is periodic
with period $T=1$ for $N$ odd only, 
while it is periodic with period $T=2$
for $N$ even; in the latter case, $D_N(k)$
changes its sign for $k \to k +1$.
\item
{\it Principal maxima at integer momenta 
$k=1,2,3,\cdots$.}

The Inside Amplitude reaches a principal
maximum when $k$ approaches an integer:
\beq
\lim_{k \to n} A_N(k) \, = \, N ;
\qquad
n \, = \, 1,2,3,\cdots.
\eeq
\item
{\it Zeroes for non-resonating momenta at zero coupling.}

For non-integer momenta of the form
\beq
\label{k_non_integer}
k \, = \, \frac{s}{N} ,
\eeq
i.e. with $s$ not an integer multiple of $N$,
\beq
s \, \notin \, N \NN_+ 
\, \equiv \, 
\big\{
N, \, 2N, \, 3N, \, \cdots
\big\},
\eeq
or, even more explicitly,
\beq
s \, = \, 1,2,3, \cdots, N-1, N+1, \cdots, 2N-1, 2N+1, \cdots,
\eeq
the Inside Amplitude exactly vanishes:
\beq
A_N\left(\frac{s}{N}\right) \, = \, 0.
\eeq
That occurs because the numerator on the r.h.s. of
the defining eq.(\ref{Dir_ker}) vanishes, while the
denominator does not.
\item
{\it Local maxima in the midpoints of non-resonating momenta for
$g \to 0$.}

By looking at fig.~\ref{fig_pinsamp},
we see that the local maxima of the Inside 
Amplitude are roughly in the middle of a pair of adjacent zeroes
and are close to the maxima of its numerator, 
namely the function $|\sin(\pi N k)|$. 
The latter lie at
\beq
\label{k_midpoints}
k \, = \, \left( u + \frac{1}{2} \right) \frac{1}{N};
\eeq
where $u$ is any signed integer,
\beq
u \, = \, \cdots, \, -2, \, -1, \, 0, \, 1, \, 2, \, 3, \, \cdots.
\eeq
Actually, by looking carefully at fig.~\ref{fig_pinsamp},
we see that the maxima of $|\sin(\pi N k)|$
closest to the point $k=1$, namely the points 
\beq
k \, = \, 1 \, \pm \, \frac{1}{2N} 
\eeq
do not correspond to any maxima of the Inside Amplitude
and have therefore to be excluded%
\footnote{
We may think that, in going from the function 
$|\sin(\pi N k)|$ to the function $A_N(k)$,
the maxima of the former, 
at $k \, = \, 1 \, \pm \, 1/2N$,
merge together to form the principal
maximum of the Inside Amplitude at $k=1$.
}.
Since $A_N(k)$ is periodic of period $T=1$, 
the above points are equivalent to the points 
\beq
k \, = \, \pm \, \frac{1}{2N}, 
\eeq
obtained by taking $u=0$ and $u=-1$ in eq.(\ref{k_midpoints})
respectively.

The above rule to find the local maxima of
the Inside Amplitude $A_N(k)$ can be analytically 
justified as follows.
In the Large-Volume or Quasi-Continuum case, $N \gg 1$,
the numerator of the Inside Amplitude, $|\sin(\pi N k)|$, 
is rapidly varying with $k$, while the denominator,
$|\sin(\pi k)|$ is slowly varying, as the factor $N$
is missing in the latter case.
That implies that, in a small momentum interval
$\delta k \ll 1$, one can consider the denominator
as a constant and vary the numerator only.
Therefore the local maxima of the Inside Amplitude
are approximately at the momenta $k$
where $\sin(\pi N k)$ is equal to $\pm 1$, 
i.e. at the points specified by eq.(\ref{k_midpoints}), 
where the index $u$ is any signed integer different 
from $0$ and $-1$.

It is convenient to re-express the above result
by explicitly taking into account the periodicity of $A_N(k)$.
By means of the euclidean division of the index $s$ by $N$,
one can write:
\beq
s \, = \, n N \, + \, h,
\eeq
where
\beq
n \, = \, \cdots, \, -2, \, -1, \, 0, \, 1, \, 2, \, 3, \, \cdots
\eeq
is the quotient and $h$, taken in the range
\beq
\frac{N}{2} \, < \, h \, \le \, \frac{N}{2};
\qquad
h \, \ne \, 0;
\eeq
is the remainder.
The non-resonating momentum in eq.(\ref{k_non_integer}) 
can then be written
\beq
k \, = \, n \, + \, \frac{h}{N}.
\eeq
A local maximum of the Inside Amplitude
is therefore at
\beq
\label{k_alla fine}
k \, \simeq \, n \, + \, \frac{h+1/2}{N};
\qquad h \, \in \, \ZZ \setminus \left\{ 0,-1 \right\}.
\eeq
On the above points, the Inside Amplitude takes 
the approximate values
\beq
\frac{ A_N \big[ k \, = \, n + (h+1/2)/N \big] }{N}
\, \simeq \, 
\frac{1}{\pi \, |h+1/2|}.
\eeq
The highest peak are
\beq
\frac{1}{\pi \, |h + 1/2|}
\, \simeq \,  
0.212, \,\, 0.127, \,\, 0.0909, \,\, 0.0707, \,\, 0.0579,
\eeq
for
\beq
h \, = \, 1, \,\, 2, \,\, 3, \,\, 4, \,\, 5
\qquad
\mathrm{or}
\qquad
h \, = \, -2, \,\, -3, \,\, -4, \,\, -5, \,\, -6,
\eeq
respectively.
As we see from the above formulae,
the highest peak after the principal one $A_N(k=\mathrm{integer})=N$
is reached for $h=1$ (or $h=-2$) 
and is roughly $21\%$ of its value.
\item
{\it Unitary value in the large-coupling limit.}

For non-integer momenta of the form
\beq
\label{large-coupling_val}
k \, = \, \frac{u}{N+1},
\eeq
with $u$ any integer not multiple of $N+1$,
\beq
u \, \in \, \ZZ \, \setminus \, (N+1) \, \ZZ,
\eeq
the Inside-Amplitude equals exactly one:
\beq
\label{Ains_large_coup}
A_N\left(\frac{u}{N+1}\right) \, = \, 1.
\eeq
The above equation follows immediately from 
\beq
D_N\left(\frac{u}{N+1}\right) \, = \, \pm \, 1.
\eeq
By means of the euclidean division of $u$ by $N+1$,
the momentum $k$ in eq.(\ref{large-coupling_val}) 
can be written 
\beq
k \, = \, n \, + \, \frac{l}{N+1}, 
\eeq
where
\beq
n \, = \, \cdots, \, -2, \, -1, \, 0, \, 1, \, 2, \, 3, \, \cdots
\quad
\mathrm{and} 
\qquad
\frac{N+1}{2} \, < \, l \, \le \, \frac{N+1}{2}.
\eeq
Eq.(\ref{Ains_large_coup}) can be rewritten as
\beq
A_N\left( n + \frac{l}{N+1} \right) \, = \, 1.
\eeq
In the large-coupling limit, the ratio
of the inside amplitude over the outside one 
goes to one, i.e. the amplitude is equal in
both cavities.
That is expected according to the fact
that the potential barrier vanishes for 
$g \to \infty$.
\end{enumerate}
%
%


\subsubsection{Phase Shift as function of momentum}

The Phase Shift $\delta_N$, as a function of the
momentum $k$, $\delta_N=\delta_N(k)$ (eq.(\ref{delta_N_def})),
has the following properties 
(see fig.~\ref{fig_pfase}):
\begin{enumerate}
\item
{\it Unitary Period.}
 
The Phase Shift is periodic of period $T=1$, for both
$N$ odd and $N$ even:
\beq
\delta_N(k+1) \, = \, \delta_N(k)
\qquad k \, \in \, \RR.
\eeq
\item
{\it Jump equal to $2 \pi$ at integer momenta.}

With the following range of the two-argument arctangent, 
\beq
- \, \pi \, < \, \arctan(x,y) \, \le \, \pi,
\eeq
i.e. with the discontinuity along the negative axis,
the phase $\delta_N(k)$ has a jump 
(i.e. a finite discontinuity)
equal to $2\pi$ whenever 
the momentum $k$ is an integer,
\beq
k \, = \, n = 1,2,3,\cdots.
\eeq
That is because the first argument of the arctangent
is negative in a neighborhood of $k=n$, while
the second one changes its sign by crossing the point 
$k=n$.
\item
{\it Jump of $\pi$ at non-resonating momenta for $g\to 0$.}

The phase $\delta_N(k)$ has a jump  
equal to $\pi$ at non-integer momentum of the form
\beq
\label{k_non_reson}
k \, = \, \frac{s}{N}; 
\eeq
i.e. when $s$ is not a multiple of $N$,
\beq
s \, \notin \, N \, \NN_+ 
\, \equiv \, 
N \left\{ 1, 2, 3, \cdots\ \right\}
\, \equiv \, 
\left\{ N, 2N, 3N, \cdots\ \right\}.
\eeq
That is because because the function $\sin(\pi N k)$,
appearing in both arguments of the arctangent
function (cfr. eq.(\ref{delta_N_def})), 
changes its sign by crossing a momentum
of the form (\ref{k_non_reson}).
\item
{\it Resonant behavior in the midpoints of non-resonant momenta 
for $g \to 0$.}

In terms of the standard arctangent function
(periodic of period $\pi$ rather than $2\pi$
as the two-argument arctangent), 
the phase shift has the simple expression
\beq
\delta_N(k) \, = \, - \, \pi (N+1) \, k \qquad (\mathrm{mod}\,\pi).
\eeq
According to the above formula, the phase passes through the resonant value $\pi/2$ (modulo $\pi$) whenever
\beq
k \, = \, n + \frac{h+1/2}{N+1} \, \simeq \, n + \frac{h+1/2}{N}
\qquad (N \gg 1).
\eeq 
For $n=1$ and $h = 1$, for example, a resonance occurs for 
$k \simeq 1 + 1.5/N$.
Therefore, for large $N$, 
the phase shift passes through $\pi/2$
(modulo $\pi$) when 
the Inside Amplitude reaches a maximum
(cfr eq.(\ref{k_alla fine})).
\end{enumerate}

%
\begin{figure}[ht]
\begin{center}
\includegraphics[width=0.5\textwidth]{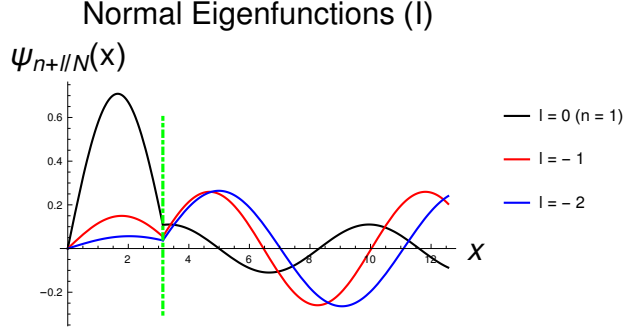}
\footnotesize
\caption{
\label{fig_peigenfun1}
\it 
plots of normal eigenfunctions $\Psi_{n+l/N}(g;x)$ 
in the case of the rather small coupling $g=0.05$
and $N=9$.
The black continuous line represents the 
first resonance, $n=1,l=0$;
the red continuous line the level right below it, 
$n=1,l=-1$;
finally, the blue continuous line is level $l=-2$.
A thick dot-dashed green vertical line indicates
the point $x=\pi$, the support of the $\delta$-function 
potential.
}
\end{center}
\end{figure}
%


\subsubsection{Normal Eigenfunctions for different couplings}

The normal eigenfunctions, as functions of the integer index 
$s=1,2,3,\cdots$ and of the coupling $g$, 
are simply obtained by substituting, 
inside eq.(\ref{old_ef}) or (\ref{new_ef}), 
the independent momentum $k$ with the general solution of 
eq.(\ref{eq_basic}):
\beq
\Psi_{s/N}(g;\, x) 
\, \equiv \, 
\psi\left[ k \mapsto k_{s/N}(g); \, x \right];
\qquad
s = 1,2,3,\cdots.
\eeq
In order to simplify the notation,
the explicit dependence on the coupling $g$ 
in the eigenfunctions will often be omitted,
so that the latter will be written simply
$\Psi_{s/N}(x)$.
%
\begin{figure}[ht]
\begin{center}
\includegraphics[width=0.5\textwidth]{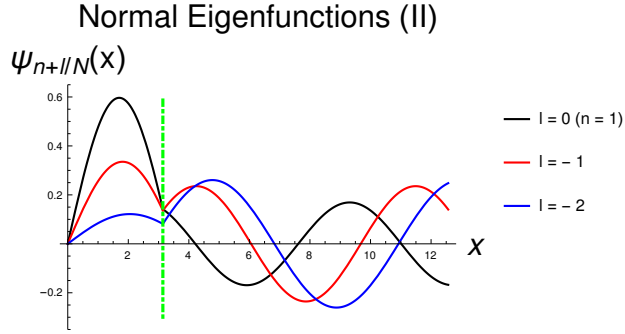}
\footnotesize
\caption{
\label{fig_peigenfun2}
\it 
plots of normal eigenfunctions $\Psi_{n+l/N}(x)$ for 
$g=0.1$ and $N=9$.
As in the previous plot,
the black continuous line is the 
first resonance, $l=0$,
the red continuous line the level $l=-1$
and the blue continuous line the level $l=-2$.
}
\end{center}
\end{figure}
%

\begin{figure}[ht]
\begin{center}
\includegraphics[width=0.5\textwidth]{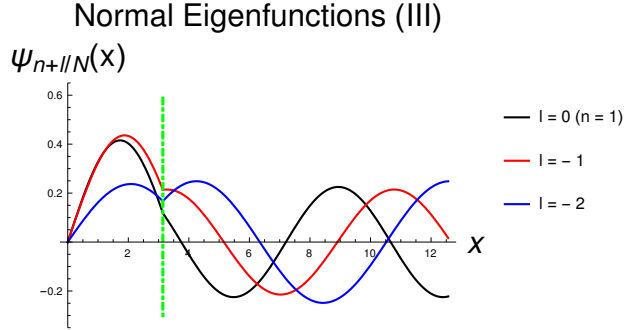}
\footnotesize
\caption{
\label{fig_peigenfun3}
\it 
plots of normal eigenfunctions $\Psi_{n+l/N}(x)$ for 
the relatively large value of the coupling $g=0.2$
and $N=9$.
As in previous plots, the black continuous line represents 
the first resonance, $n=1,l=0$,
the red continuous line the level $l=-1$
and the blue continuous line the level $l=-2$.
}
\end{center}
\end{figure}

In figs.$\,$\ref{fig_peigenfun1}$-$\ref{fig_peigenfun3}
we plot the eigenfunctions
of the first resonance, $n=1$, $l=0$,
and of the two levels right below it, $l=-1$ and $l=-2$, 
for different values of the coupling $g$ and
for $N=9$.
In fig.~\ref{fig_peigenfun1} the coupling chosen
is rather small: $g=0.05$;
the effective coupling $gN = 0.45 < 1$ (see later).
The $l=0$ eigenfunction (the black line) exhibits a rather
marked resonance behavior, as its amplitude inside the
small cavity is $6\div 7$ larger than outside it 
(i.e. inside the large cavity);
furthermore, the $l=0$ eigenfunction is by far the 
largest one inside the small cavity.
The $l=-1$ eigenfunction (the red line) is smaller roughly by
a factor two inside the small cavity than outside it,
not showing therefore any resonant behavior.
Note the (finite) discontinuity of the first derivatives
of all the eigenfunctions at $x=\pi$.
%

In fig.~\ref{fig_peigenfun2} we have taken $g=0.1$, 
i.e. we have increased the coupling by 
a factor two with respect to the previous plot;
the effective coupling $gN=0.9$, i.e. it
is close to unity.
The $l=0$ eigenfunction has a less marked
resonance behavior compared to the
previous plot, as it is larger 
by a $3 \div 4$ factor only inside the
small cavity than outside it;
however, it is still the largest one inside the 
small cavity.
Contrary to previous plot,
the $l=-1$ eigenfunction is (slightly) larger
inside the small cavity than outside it,
showing therefore a mild resonant behavior.
As in the previous plot, the $l=-2$ eigenfunction 
is still substantially smaller inside the small
cavity than outside it, indicating
the absence of any resonating behavior.

In fig.~\ref{fig_peigenfun3} the coupling 
has been increased by a further factor two:
$g=0.2$; 
the effective coupling $gN = 1.8$,
i.e. it is substantially larger than unity.
There is still an overall, though modest,
resonant behavior.
The main different with respect to
the previous plots is that 
the $l=-1$ eigenfunction is (a bit)
larger than the resonant one ($l=0$)
inside the small cavity;
the resonant behavior is "transferred" 
from the level $l=0$ to the level $l=-1$.
Furthermore, contrary to
previous plots, the $l=-2$ eigenfunction
is (slightly) larger inside the small
cavity than outside it.


\subsubsection{Inside Amplitude as function of the coupling}

In order to understand the resonance phenomena occurring 
in the spectrum of finite-volume Winter model, 
let us consider in this section 
the Inside Amplitude $A_N$ as a function of the coupling 
$g$, which is our "tuning" or "detuning" parameter.
The desired function is simply obtained by means of the replacement
\beq
k \, \mapsto \, k_{s/N}(g)
\eeq
inside the function $A_N(k)$, giving rise to
\beq
A_{s/N}(g) \, \equiv \,  A_N\left[ k_{s/N}(g) \right];
\qquad
s = 1,2,3,\cdots; \quad k > 0.
\eeq
%
\begin{figure}[ht]
\begin{center}
\includegraphics[width=0.5\textwidth]{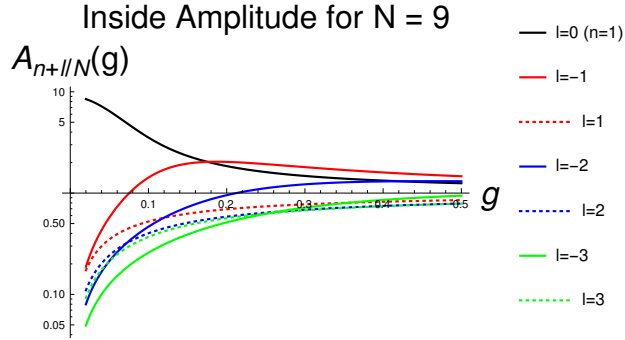}
\footnotesize
\caption{
\label{fig_pinsideamp}
\it Inside Amplitude $A$ as a function of the coupling $g$
for various levels around and including the
first resonance, $A = A_{n+l/N}(g)$, in the weak-coupling region
$0.02 \le g \le 0.5$ (which is the relevant one in the study of 
resonances) for $N=9$.
The vertical scale --- that of the Amplitude --- 
is logarithmic, while the horizontal one
--- that of the coupling --- is linear.
}
\end{center}
\end{figure}
%
In fig.~\ref{fig_pinsideamp}
the Inside Amplitude $A_{n+l/N}(g)$
for the first resonance, $n=1,\,l=0$,
is plotted as a black continuous line.
We see that $A_1(g)$ is the largest amplitude 
in the interval ranging
from the free-theory point $g=0$, 
where 
\beq
A_1(g=0) \, = \, N \, = \, 9 \, \gg \, 1, 
\eeq
up to the critical value of the coupling
\beq
g \, = \, \bar{g}_{cr} \, = \, 0.176 
\, \approx \, \frac{3}{2N} \, = \, 0.167, 
\eeq
where it is surpassed by the $l=-1$ level (the red continuous line). At the critical point both Amplitudes have 
roughly a factor two enhancement, 
\beq
A_1\left(\bar{g}_{cr}\right) 
\, = \, 
A_{1-1/N}\left(\bar{g}_{cr}\right) 
\, = \, 2.03.
\eeq
As already seen in the previous section, 
by increasing the coupling above $\bar{g}_{cr}$,
the resonant behavior is transferred from the level $l=0$
down to the level right below it, namely the level $l=-1$.
Actually, the $l=-1$ Inside Amplitude has a maximum at 
a value of the coupling a bit larger than $\bar{g}_{cr}$,
namely at
\beq
\bar{g}_{l=-1} \, = \, 0.187,
\eeq
where it is slightly greater than at $\bar{g}_{cr}$:
\beq
A_{1-1/N}\left( \bar{g}_{l=-1} \right) \, = \, 2.04.
\eeq
The $l=-2$ Inside Amplitude (the continuous blue line) 
has a rather flat maximum at 
\beq
\bar{g}_{l=-2} \, = \, 0.461, 
\eeq
where it has a quite modest $30\%$ enhancement:
\beq
A_{1-2/N}\left( \bar{g}_{l=-2} \right) \, = \, 1.31.
\eeq
Finally, the $l=-3$ Inside Amplitude (the continuous green line)
has a really flat maximum at 
\beq
\bar{g}_{l=-3} \, = \, 1.267, 
\eeq
where it has basically no enhancement,
\beq
A_{1-3/N}\left(\bar{g}_{l=-3}\right) \, = \, 1.07.
\eeq
Due to the large values of $\bar{g}_l$
found in the cases $l=-2$ and $l=-3$,
one can reasonably state that these maxima 
cannot be described in perturbation theory. 
Finally, let us note that the Inside Amplitudes of the levels 
with $l>0$ (the dotted lines of various colors) 
never become larger than one,
in complete agreement with the fact that they never 
resonate in the repulsive case $(g>0)$.

%
\begin{figure}[ht]
\begin{center}
\includegraphics[width=0.5\textwidth]{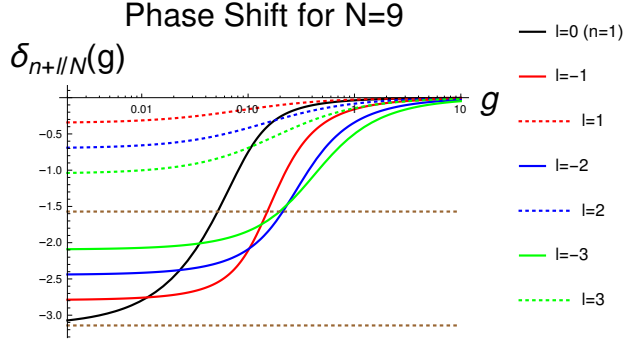}
\footnotesize
\caption{
\label{fig_prelphase}
\it Phase Shift $\delta$ of the outside amplitude
with respect to the inside one
as a function of the coupling $g$,
$\delta = \delta_{s/N}(g)$,
for the levels close to the first resonance 
($s=N$ or, equivalently $n=1$ and $l=0$),
for a wide range of $g$, for $N=9$. 
The horizontal dotted brown lines
denote the phases $-\pi/2$ and $-\pi$.
The horizontal scale
(that of the coupling $g$)
is logarithmic while the vertical scale
(that of the phase $\delta$)
is linear.
}
\end{center}
\end{figure}
%


\subsubsection{Phase Shift as a function of the coupling}

\begin{figure}[ht]
\begin{center}
\includegraphics[width=0.5\textwidth]{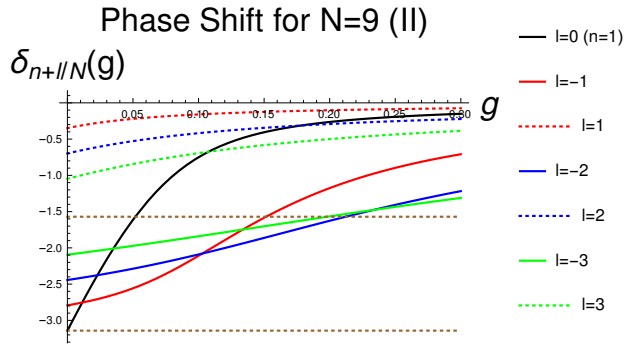}
\footnotesize
\caption{
\label{fig_prelphase2}
\it 
\it Phase Shift $\delta = \delta_{s/N}(g)$,
for the levels close to the first resonance ($n=1,\,l=0$),
for small values of $g$ and for $N=9$.  
In this plot both scales are linear.
}
\end{center}
\end{figure}

In this section we study the phase shift $\delta$
as a function of the coupling $g$,
\beq
\delta_{s/N}(g) 
\, \equiv \,  
\delta_N\left[ k_{s/N}(g) \right];
\eeq
where
\beq
s = 1, 2, 3, \cdots, N-1, N, N+1, \cdots; 
\qquad 
k \, \in \, \RR^+.
\eeq
We consider the levels close to the first resonance
both in figs.$\,$\ref{fig_prelphase} and \ref{fig_prelphase2}.
As well known from scattering theory, 
an eigenfunction with a momentum $k = s/N + \mathcal{O}(g)$ 
exhibits a resonant behavior 
at a value of the coupling $g_{res}$
where its phase passes 
through $\pi/2$ (modulo $\pi$),
\beq
\delta_{s/M}\left(g = g_{res}\right) \, = \, \frac{\pi}{2}
\qquad (\mathrm{mod} \, \pi),
\eeq
with a large first derivative, 
\beq
\frac{d\delta_{s/M}}{dg}\left(g_{res}\right) \, \gg \, 1.
\eeq
By looking at figs.$\,$\ref{fig_prelphase} and \ref{fig_prelphase2},
we see that the continuous black curve, representing the phase shift 
of the first resonance, $n=1$, $l=0$, 
crosses the upper brown dotted line $-\pi/2$,
i.e. resonates, at 
\beq
g \, = \, \hat{g}_{l=0} \, = \, 0.0518 
\, \approx \, \frac{1}{2N} \, = \, 0.556,
\eeq
where the Inside Amplitude is large
and $\mathcal{O}(N)$,
\beq 
A_1\left(\hat{g}_{l=0}\right) \, = \, 6.31.
\eeq
The continuous red curve $l=-1$  intersects 
the brown line $-\pi/2$, i.e. resonates, at
\beq
\hat{g}_{l=-1} \, = \, 0.152 \, \approx \, 
\frac{3}{2N} \, = \, 0.167, 
\eeq
a coupling value which is not far
from the peak of the corresponding 
Inside Amplitude, 
$\bar{g}_{l=-1} \, = \, 0.187$.
On the contrary, the continuous blue line $l=-2$ 
intersects the upper brown line at 
\beq
g \, = \, \hat{g}_{l=-2} \, = \, 0.212,
\eeq
a value of the coupling quite far from the maximum of the 
corresponding Inside Amplitude, $\bar{g}_{l=-2} = 0.461$.
Finally, the continuous green line, 
representing the Phase Shift of the level $l=-3$, 
crosses the brown line close to the intersection 
point of the $l=-2$ level.
This sort of approximate degeneracy may be interpreted
by saying that, by taking $l \le -2$, we enter,
as far as resonance dynamics is concerned,
a strong-coupling region,
where usual perturbation theory does not 
work (the effective coupling $N \hat{g}_{l=-2} \approx 2$).
In order to study these $l$ values in
perturbation theory, one has to go to larger values 
of $N$.

%
%

The Phase Shifts of the levels with $l>0$
(the dotted lines of various colors) have a very 
small variation with $g$ and, in particular, 
they never cross the brown line $-\pi/2$, implying the absence 
of any resonant behavior for $g>0$, in agreement
with the results obtained for the Inside Amplitude
(see previous section).


\section{Ordinary Perturbation Theory}
\label{ord_pert_th}

In the analytic calculations, we always assume 
\beq
0 \, < \, g \, \ll \, 1, 
\eeq
i.e. a weakly-repulsive interaction, in order
to use perturbation theory in $g$ and avoid
bound-state effects. 
The momentum on the r.h.s. of eq.(\ref{order_zero})
can be considered the zero order term
of a perturbative expansion in the coupling $g$.
One has to distinguish between: 
\begin{enumerate}
\item
{\it Resonant case,} when $s$ is an integer 
multiple of $N$,
\beq
s \, = \, n \, N;
\qquad n = 1, 2, 3, \cdots,
\eeq
so that the unperturbed momentum is integer,
\beq
k^{(0)} \, = \, n;
\eeq
\item
{\it Non-resonant case,} when $s$ is not multiple of $N$,
i.e. the fraction on the r.h.s. of eq.(\ref{order_zero})
is not apparent;
the unperturbed momentum is written in this case
\beq
\label{k_0_nonres_2}
k^{(0)} \, = \, n \, + \, \frac{l}{N};
\eeq
with
\beq
- \, \frac{N}{2} \, < \, l \, \le \, \frac{N}{2};
\qquad
l \, \ne \, 0.
\eeq
\end{enumerate}
The perturbative expansion of the constant levels $p_n \equiv n$
is of course trivial, as these levels do not depend on $g$.


\subsection{Resonant case}

In the resonant case, the expansion is written
\beq
\label{k_reson_nonres}
\qquad\qquad
k_n \, = \, k_n(g) \, = \, 
n \left( 1 \, + \,  \sum_{i=1}^\infty g^i \, c_n^{(i)} \right);
\qquad
n = 1, 2, 3, \cdots;
\eeq
As already discussed, we label the momentum levels by means of their 
limiting values for $g \to 0$:
\beq
k_n(g=0) \, = \, n.
\eeq
By recursively solving in $g$ the 
momentum eigenvalue eq.(\ref{eq_basic}),
one obtains for the first few coefficients:
\bea
\label{coef_ris_straight}	
c^{(1)}_n &=& - \left( 1 \, + \, \frac{1}{N} \right);
\\
c^{(2)}_n &=& + \left(1 \, + \, \frac{1}{N}\right)^2 ;
\nonumber\\
c^{(3)}_n &=& + \,
\left( 1 + \frac{1}{N} \right)^3
\left( \frac{\,\pi^2}{3} \, n^2 N \, - \, 1 \right) .
\eea
Let us comment on the above results.
The resonances of the usual, infinite-volume model
are related to the poles in the complex $k$-plane
\cite{primo,winter}
\beq
\label{poles_Infinite_Volume}
k \, = \, \chi_n(g) \, = \, 
n \Big( 1 \, + \,  g \, b^{(1)}_n
\, + \,  g^2 \, b^{(2)}_n \, + \, \cdots \Big);
\eeq
where
\beq
b^{(1)}_n \, = \, - \, 1;
\qquad
b^{(2)}_n \, = \, 1 \, - \, i \pi n.
\eeq
For $N\to \infty$, i.e. in the infinite-volume limit,
the first-order coefficient $c^{(1)}_n=c^{(1)}_n(N)$
smoothly goes into the corresponding 
coefficient $b^{(1)}_n$ of the infinite-volume theory,
\beq
\lim_{N \to \infty} c^{(1)}_n \, = \, b^{(1)}_n ;
\eeq
furthermore, $c^{(2)}_n$ goes into the real part of 
$b^{(2)}_n$,
\beq
\lim_{N \to \infty} c^{(2)}_n \, = \, 
\mathrm{Re} \, b^{(2)}_n ;
\eeq
Surprises come at third order:
by looking at the expression of the 
third-order coefficient $c^{(3)}_n$ above,
we find a secular term proportional to $N$,
giving a contribution to $k_n(g)$ of the form
$g^3 N$.
By analyzing higher-order coefficients 
(which we have not displayed),
we find that the occurrence of positive
powers of $N$ in the coefficients $c^{(i)}_n$ 
is a general phenomenon:
The term $g^i c_n^{(i)}$, 
for $i = 3, 4, 5, \cdots$,
contains monomials of the form
\beq
\label{secular_terms}
g^2 (g N)^{i-2}, \quad g^3 (g N)^{i-3},
\quad g^4 (g N)^{i-4}, \quad \cdots .
\eeq
To have a convergent series,
we have therefore to restrict 
the range of the perturbative expansion
to the parameter-space region
\beq
\label{spoil_PT}
0 \, < \, g \, N \, \ll \, 1,
\qquad
\mathrm{or\,\, equivalently}
\qquad
0 \, < \, g \, \ll \, \frac{1}{N} .
\eeq
For small sizes of the larger cavity,
let's say $N = \mathcal{O}(1)$, 
the secular terms shown in (\ref{secular_terms})
do not present any problem, as they do not
lead to any substantial enhancement
in the coefficients $c_n^{(i)} = c_n^{(i)}(N)$
entering the expansion of $k_n = k_n(g,N)$.
On the contrary, for large $N$, the terms in (\ref{secular_terms})
tend instead to spoil the convergence of the perturbative expansion.
The occurrence of such secular terms
in the momentum expansion 
implies in particular that we can no more 
take the infinite-volume limit $N \to \infty$
in a straightforward way, as we have
made in the first-order and second-order
cases.

In the very-small coupling region 
specified by the inequalities in (\ref{spoil_PT}), 
the allowed momentum
\beq
\label{doublet_1}
k_n(g) \, = \, 
n \, - \, g \, n \left( 1 + \frac{1}{N} \right) \, + \, \cdots
\eeq
is very close to --- slightly below ---
the exceptional, exactly-integer momentum
\beq
\label{doublet_2}
p_n \, \equiv \, n,
\eeq
whose corresponding eigenfunction has been given in 
eq.(\ref{eigen_k_integer}).
Their separation vanishes indeed in the free limit:
\beq
p_n \, - \, k_n(g) \, = \, 
g \, n \left( 1 + \frac{1}{N} \right) \, + \, \cdots.
\eeq
The conclusion is that for very small couplings, $g\ll 1/N$,
the quasi-degenerate doublet 
\beq
\big( k_n(g), \, p_n \big)
\eeq
corresponds to the $n^{\mathrm th}$ resonance of the
usual, infinite-volume model, for any $n=1,2,3,\cdots$.
Ordinary perturbation theory therefore allows
an analytic description of the exact double degeneracy for $g \to 0$,
which we have observed in the plots of the exact levels
in figs.$\,$\ref{fig_pNsmall}$-$\ref{fig_pNlarge} 
and in fig.~\ref{fig_pdifferenza}. 
%
\begin{figure}[ht]
\begin{center}
\includegraphics[width=0.5\textwidth]{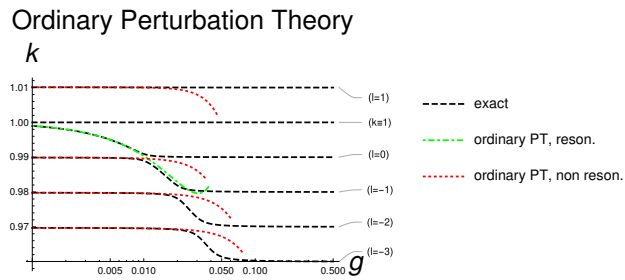}
\footnotesize
\caption{
\label{fig_pordpert}
\it momentum levels around the first or fundamental 
resonance $(n=1)$ as functions of the coupling, $k=k(g)$,
in the large-volume case $M=100$.
The black dashed lines are the plots of the exact levels,
the green dot-dashed line represents
the ordinary perturbative formula for the resonant level ($l = 0$),
while the red lines are the plots of the
perturbative formulae in the  non-resonant cases ($l \ne 0$).
The horizontal scale is logarithmic.
}
\end{center}
\end{figure}
%


\subsection{Non-resonant case}

In the non-resonant case, the perturbative expansion
of the momenta, according to eq.(\ref{k_0_nonres}), is written:
\beq
\qquad
\qquad
\qquad
k \, = \, k_{n+l/N}(g) \, = \, 
\left( n \, + \, \frac{l}{N} \right)
\left(
1 \, + \, \sum_{i=1}^\infty
g^i \, c^{\,(i)}_{n + l/N}
\right)
\qquad
\qquad
(l \ne 0).
\eeq	
As in the resonant case, the momentum levels,
as functions of the coupling, are labeled 
by their limiting values for $g\to 0$:
\beq
k_{n+l/N}(g=0) \, = \, n \, + \, \frac{l}{N}.
\eeq
The lowest-order coefficients of the perturbative expansion 
have the following explicit expressions:
\bea
c_{n + l/N}^{\,(1)} &=& - \, \frac{1}{N};
\\
c_{n + l/N}^{\,(2)} &=& 
+ \, \frac{\pi}{N} 
\left(n + \frac{l}{N}\right) \cot\left(\frac{\pi l}{N}\right)
+
\frac{1}{N^2};
\nonumber\\
c_{n + l/N}^{\,(3)} &=& 
- \, \frac{\pi^2}{N} \left( 1 - \frac{1}{N} \right) 
\left(n + \frac{l}{N}\right)^2 \csc^2\left(\frac{\pi l}{N}\right) \, +
\nonumber\\
&& - \, \frac{3 \pi}{N^2} \left(n + \frac{l}{N}\right) 
\cot\left(\frac{\pi l}{N}\right)
+ \frac{4 \pi^2}{3N} \left(n + \frac{l}{N}\right)^2
- \frac{1}{N^3};
\nonumber
\eea
where $\cot(x)$ and $\csc(x)$ are the 
cotangent and the cosecant of $x$ respectively.
By expanding the above coefficients for large $N$,
one obtains: 
\bea
\label{fix_ord_no_reson}
c_{n+l/N}^{\,(1)} &=& - \, \frac{1}{N};
\nonumber\\
c_{n+l/N}^{\,(2)} &=& + \, \frac{n}{l} 
\, + \, \frac{1}{N}
\, + \, \mathcal{O}\left(\frac{1}{N^2}\right);
\nonumber\\
c_{n+l/N}^{\,(3)} &=& - \, \left(\frac{n}{l}\right)^2 \, N
\, + \, \frac{n\,(n-2l)}{l^2} 
\, + \, \frac{\pi^2 \, n^2 - n/l-1}{N}
\, + \, \mathcal{O}\left(\frac{1}{N^2}\right). 
\eea
In general, any additional power of the coupling $g$
brings, inside its corresponding coefficient, an additional
power of $N$;
The third-order coefficient, in particular, has a
secular term linear in $N$, as in the resonant case.


\subsection{Discussion}

In fig.~\ref{fig_pordpert} we plot the exact momentum 
levels around the first resonance $(n=1)$ together
with their ordinary perturbative expansions, for $M=100$.
The dot-dashed green curve, representing the perturbative expansion 
of the first resonant level 
(the plot of eq.(\ref{k_reson_nonres}) with $n=1$
truncated at third order in $g$)
is very close to the corresponding exact curve (the black dashed line) 
from $g=0$ up to $g \approx 10^{-2}$; 
around this point 
(as already seen in the previous $M=200$ plot, fig.~\ref{fig_pNlarge}), 
the exact curve basically 
reaches its lower asymptote, showing a quick and large 
variation of its first derivative and getting pretty 
close to the lower, non-resonating, curve $l=-1$.
For values of the coupling in the region 
\beq
10^{-2} \, \,\lsim \,\, g \,\, \lsim \,\, 2 \times 10^{-2},
\eeq
the resonant perturbative curve is rather close to 
(a little above actually) the exact curve of the non-resonant 
level $l=-1$.
Above $g \approx 2 \times 10^{-2}$, the
perturbative curve shows an unphysical rise
and is no more able to describe any level.
It is noticeable that a single perturbative formula
(approximately) describes {\it two} distinct levels
--- rather than one ---
in two different coupling regions.

The plots of the perturbative expansions
of the non-resonant levels for $l<0$ 
(the red dotted lines below the green dot-dashed line)
remain close to the corresponding exact curves
(black dashed lines)
from $g=0$ up to the coupling region where 
the latter make a transition from the upper asymptote 
to the lower one.
The perturbative curves are not able to correctly
reproduce this transition, showing some sort of "delay".
As in previous fig.~\ref{fig_pNlarge},
the exact curve of the non-resonant level $l=1$
(the level right above $k \equiv 1$)
is pretty flat in all the coupling range;
the corresponding perturbative curve
(the red dotted line above the green dot-dashed line)
is close to the exact one from $g=0$
up to approximately $g \approx 2 \times 10^{-2}$,
where it shows an unphysical decrease.


\section{Resummed Perturbation Theory}
\label{res_pert_th}

The parameter-space region
\beq
g \, N \, \gsim \, 1,
\eeq
which is out of control with ordinary
perturbation theory as we have seen
in the previous section,
can be studied by means of improved
perturbative expansions, 
generated in the limit
\beq
\label{lim_resum}
g \, \to \, 0^+, \quad N \, \to \, + \, \infty
\qquad \mathrm{with} \qquad
\xi \, \equiv \, g \, N \, \to \, \mathrm{constant} .
\eeq
As in the case of ordinary perturbation theory,
we have to treat separately the resonant momenta 
and the non-resonant ones.


\subsection{Resonant case}

In the resonant case, the $k_n$'s
are represented by function series of the form:
\beq
\label{k_ris_res}
k_n \, = \, n 
\left(1 \, + \, \sum_{i=1}^\infty g^i \, h_n^{(i)}(\xi) \right).
\eeq
By means of standard multi-scale techniques \cite{nayfeh},
one obtains for the lowest-order functions:
\bea
h_n^{(1)}(\xi) &=& - \, 1;
\nonumber\\
h_n^{(2)}(\xi) 
&=& 
- \, \nu \, \cot\left( \nu \, \xi \right) \, + \, 1 ;
\\
h_n^{(3)}(\xi)
&=&
+  \, \nu^3 \xi \cot^3\left( \nu \, \xi \right)
\, - \, \nu^2 (1+\xi) \, \cot^2\left( \nu \, \xi \right)
\, + \, 
\nonumber\\
&& + \, \nu \left(\nu^2 \xi + 3\right) \cot\left( \nu \, \xi \right)
\, + \, \nu^2\left(\frac{1}{3} - \xi \right) \, - \, 1;
\nonumber
\eea
where we have defined 
\beq
\nu \, \equiv \, \pi \, n.
\eeq
The functions $g^i h^{(i)}_n(\xi)$, $i>1$,
have pole singularities whenever
\beq
\label{singul}
n N g \, = \, j \, = \, 1,2,3, \cdots .
\eeq
For a given size of the large cavity, i.e. for given $N$, 
and for a given resonance, i.e. for given $n$,
the functions $h^{(i)}_n(\xi)$
become singular when the coupling $g$
approaches one point of the sequence
\beq
\label{def_gj}
g_j \, \equiv \, \frac{j}{n \, N};
\qquad
j \, = \, 1,2,3, \cdots.
\eeq
As we have seen in figs.$\,$\ref{fig_pNlarge}$-$\ref{fig_pdifferenza},
the points $g_j$ are the particular values of
the coupling $g$ where the exact levels
present a large variation of their first derivatives
and quasi-degenerate doublets occur in the spectrum.

A physical argument about the origin of
the $g_j$ singularities is the following.
The eigenfunctions of a free particle in the
$[0,\pi]$ box have exactly integer momenta:
\beq
k_{\,[0,\pi]\,\mathrm{box}} \, = \, n \, = \, 1,2,3,\cdots.
\eeq	
As already discussed,
by weakly-coupling such states to a continuum,
the box becomes a resonant cavity 
and the above box momenta are subjected to
a finite renormalization \cite{primo,secondo} which, 
to first order in $g$, gives
the resonant momenta (cfr. eq.(\ref{poles_Infinite_Volume}))
\beq
\label{chi_n_fo}
\chi_n(g) \, \approx \, n \, - \, n g.
\eeq
Now, if we couple the $[0,\pi]$ box 
to the large $[\pi,L]$ box, 
rather than to the half line $[\pi,\infty)$,
we expect singularities to occur in
the perturbative expansion of $k_n = k_n(g,\xi)$, 
whenever a resonant momentum $\chi_n(g)$ of the 
infinite-volume theory becomes equal to an allowed momentum
of the particle in the large box, namely when
\beq
\chi_n(g) \, \approx \,
n \, + \, \frac{l}{N}.
\eeq 
By replacing the explicit form of $\chi_n(g)$
given in eq.(\ref{chi_n_fo}), we obtain
the relation
\beq
n N g \, \approx \, - \, l.
\eeq
If we now identify $j$ with $-l$,
we obtain eq.(\ref{singul}).
%
\begin{figure}[ht]
\begin{center}
\includegraphics[width=0.5\textwidth]{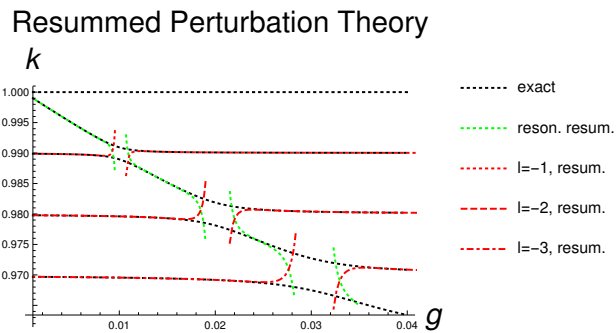}
\footnotesize
\caption{
\label{fig_presum}
\it few momentum levels below the first or fundamental 
resonance $(n=1)$ for $M=100$.
The black dotted lines are the plots of the exact levels,
the green dotted line represents the resummed
formula in the resonant case $(l=0)$,
while the red lines represent the resummed formulae in the
non-resonant cases $(l \ne 0)$. 
To be more specific, the red dotted line is the level 
$l=-1$, the red dashed line is the level $l=-2$
and the red dot-dashed line the level $l=-3$.
The singularities of the resummed formulae at the points
$g = g_j \equiv j/(nN) = j/99 \cong 0.01 \times j$,
corresponding to quasi-degenerate doublets
and large variations of their first derivatives, 
are clearly visible for $j=1,2,3$.
}
\end{center}
\end{figure}


\subsection{Non-resonant case}

In the non-resonant case, the resummed expansion
in the effective coupling $\xi$ for the quantized momentum 
$k_{n+l/N} \, = \, k_{n+l/N}(g,\xi)$ is written:
\beq
\label{resum_nores}
\qquad
\qquad
\qquad
\qquad
k_{n+l/N} \, = \, \left( n \, + \, \frac{l}{N} \right) 
\left(
1 \, + \, \sum_{i=2}^\infty g^i \, h^{(i)}_{n+l/N}(\xi)
\right)
\qquad
\qquad
(l \ne 0);
\eeq
where:
\bea
h^{(2)}_{n+l/N}(\xi) &=&
\frac{n}{l + n \, \xi} \, - \, \frac{1}{\xi};
\nonumber\\
h^{(3)}_{n+l/N}(\xi) &=& 
\frac{l \, n^2}{(l + n \, \xi)^3}
\, - \, \frac{l n}{(l + n \, \xi)^2}
\, - \, \frac{n}{l + n \, \xi}
\, + \, \frac{1}{\xi}.
\eea
%
\begin{figure}[ht]
\begin{center}
\includegraphics[width=0.5\textwidth]{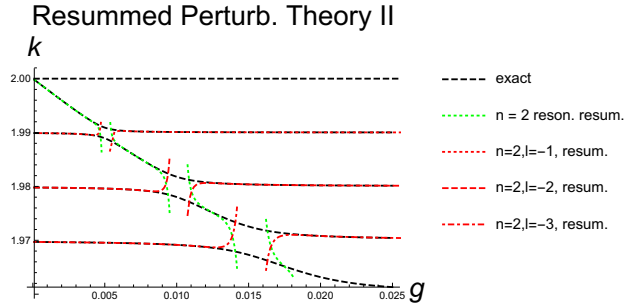}
\footnotesize
\caption{
\label{fig_presum2}
\it few momentum levels below the second or first-excited 
resonance $(n=2)$ for the case $M=100$.
The plot scheme is the same as in previous figure 
\ref{fig_presum}, so we do not repeat its description.
The singularities of the resummed formulae at 
$g  = j/198 \simeq 0.005 \times j$,
corresponding to quasi-degenerate doublets
and large variations of their first derivative, are 
clearly visible for $j=1,2,3$.
}
\end{center}
\end{figure}
%
The series begins at second order in $g$
because the term 
\beq
\frac{l}{N} \, = \, \frac{g \, l}{\xi}
\eeq
is already of first order and the perturbative expansion 
is supposed to be a small correction to the unperturbed 
result.

The momentum $k_{n+l/N}(g)$,
as given by the r.h.s. of eq.(\ref{resum_nores})
truncated at some order in $g$,
is approximated by a sum of rational functions of 
the coupling $g$,
i.e. at the end by a rational function of $g$, rather than by a polynomial
in $g$, as in ordinary perturbation theory.

For a given size of the large cavity, i.e. for given $N$, 
and for a given state, i.e. for given $n$ and $l$,
the functions $h^{(i)}_{n+l/N}$, describing the levels close
to the $n^{\mathrm{th}}$ resonance,
become singular when the coupling $g$
approaches the value
\beq
\label{l_singul}
g \, = \, g_{-l} \, \equiv \, \frac{- \,l}{n \, N};
\qquad l \, = \, \pm 1, \, \pm 2, \, \pm 3, \, \cdots.
\eeq
In the repulsive case to which we restrict, $g>0$,
the above singularity occurs for $l<0$ only, 
i.e. in the levels below the resonant one.


\subsection{Discussion}

In fig.~\ref{fig_presum} we compare the exact momentum levels
below and including the first resonance $n=1$
(black dotted lines), with the corresponding
resummed expansions (green dotted line and red lines).

Let's discuss first the simpler non-resonant case.
The resummed formula for the $l^{\mathrm th}$  level
given in eq.(\ref{resum_nores}) 
(the red dotted, dashed and dot-dashed line for $l=-1,-2$
and $-3$
respectively) correctly reproduces the corresponding 
$l^{\mathrm{th}}$ level from $g=0$
almost up to its singularity, at
\beq
g_{|l|} \, = \, \frac{|l|}{99}
\, \cong \, 0.01 \times |l|.
\eeq
For example, the $l=-1$ resummed formula (the red dotted line)
describes pretty well this level from $g=0$ almost up to its
singularity, at $g_1 \cong 0.01$.  
But what happens for larger, still perturbative, couplings?
The problem is that the resummed formula in eq.(\ref{resum_nores})
{\it is not continuous} at $g = g_{|l|}$,
because it has a pole at this point. 
It is therefore  not clear
{\it a priori} which level this formula 
is going to describe beyond this singularity, if any.
In particular, there is no reason to expect that 
this formula will still describe the $l^{\mathrm{th}}$ 
level after the singularity, i.e. for $g > g_{|l|}$.
By looking at fig.~\ref{fig_presum},
it is found that for couplings slightly above $g_{|l|}$
up to relatively large values of $g$, 
the $l^{\mathrm{th}}$ resummed curve 
reproduces quite accurately the next upper 
level, i.e. the level with index $l+1$. 
For example, if we fix the value $l=-1$ 
in eq.(\ref{resum_nores}), 
we obtain a formula which reproduces the 
exact curve $l=-1$ below the singularity, 
at $g = g_1 \cong 0.01$, 
and the resonant level $l=0$ above $g_1$.
In general, therefore, a resummed formula
for a non-resonant level describes two different 
momentum levels, in two different coupling regions.

Let's now consider the effects of the resummation
to all orders in $\xi$ of the perturbative
series for a resonant level.
The (truncated) function series on the r.h.s. of eq.(\ref{k_ris_res}) 
has a sequence of pole singularities located, in the case of the
first resonance $n=1$ and volume $N=99$, at
\beq
\qquad
\qquad
\qquad
\qquad
g \, = \, g_j \, \equiv \, \frac{j}{n\,N} \, = \, \frac{j}{99}
\, \cong \, 0.01 \times j;
\qquad j = 1, 2, 3, \cdots; \quad j \ll N.
\eeq
As in the non-resonant case,
the resummed formula in eq.(\ref{k_ris_res})
does not need to describe above each singularity, 
i.e. for $g>g_j$,
the same level described below it, i.e. for $g<g_j$.
By looking at fig.~\ref{fig_presum}, it is found
that the resummed formula for the first resonance 
(represented by the green dotted curve) reproduces pretty well, 
from $g=0$ up to almost its first singularity, 
at $g_1 \simeq 0.01$, the exact level.
Between the first and the second singularity,
i.e. for values of the coupling $g$ in the interval
\beq
g_1 \, \simeq \, 0.01 
\,\, \lsim \,\, g \,\, \lsim \,\,  
g_2 \, \simeq \, 0.02,
\eeq
the $n=1$ resummed formula describes the transition
region of the level right below the resonant one,
i.e. the non-resonant level with $l=-1$.
In general, in the coupling region
\beq
g_j \,\, \lsim \,\, g \,\, \lsim \,\, g_{j+1},
\eeq
the resummed formula for a resonant level describes 
the transition region of the non-resonant level with $l=-j$
(so long as we are in the weak-coupling regime
of course).
We find truly remarkable the fact that the resummed formula
for the $n^{\mathrm th}$ resonance describes 
the $n^{\mathrm th}$ resonance level from $g=0$
almost up to $g_1$, i.e. basically its transition region,
together with an entire sequence of non-resonant levels, 
in their respective transition regions.

In fig.~\ref{fig_presum2} we compare the 
exact and resummed momentum levels below
and including the second or first-excited resonance $n=2$.
In agreement with the general formula in eq.(\ref{def_gj}), 
the resummed formulae are singular at
\beq
g \, = \, g_j \, = \, \frac{j}{2N} \, = \, \frac{j}{198};
\qquad
j = 1,2,3,\cdots
\eeq
As we have found in fig.~\ref{fig_presum} 
for the first resonance
(as well as in the previous $M=200$ plots,
cfr. fig.~\ref{fig_pNlarge}),
the exact levels show at the points $g_j$
quasi-degenerate doublets and large
variations of their first derivatives.
In general, the plots for $n=2$ are very similar to the
corresponding ones for the first resonance,
so we do not repeat the discussion.

\begin{figure}[ht]
\begin{center}
\includegraphics[width=0.5\textwidth]{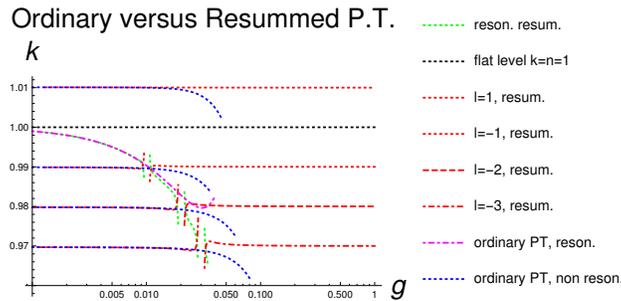}
\footnotesize
\caption{
\label{fig_pordresum}
\it ordinary versus resummed perturbation-theory
levels around the first resonance $n=1$.
The magenta dot-dashed curve is the plot the
resonant level computed in ordinary perturbation theory,
while the dotted green line describes the same
level in resummed perturbation theory.
The red dotted, dashed and dot-dashed curves 
describe the non-resonant levels in resummed 
perturbation theory for $l=\pm 1$, $l=-2$ and $l=-3$ 
respectively, while the dotted blue lines describe 
the same levels in ordinary perturbation theory.
The flat level $p_1\equiv 1$ (black dotted line),
separating the non-resonating levels (above it) 
from the resonating ones (below it), is also
shown for completeness.
The horizontal scale is logarithmic.
}
\end{center}
\end{figure}

In fig.~\ref{fig_pordresum} we compare 
the momentum levels around the first resonance ($n=1$) 
computed with ordinary perturbation theory
with the levels computed with resummed
perturbation theory.
In general, the resummed formulae work definitively 
better than the corresponding fixed-order ones. 
After excluding proper neighborhoods of the singularities 
at $g=g_j$ $(j=1,2,3,\dots)$ and proper matching,
resummed perturbation theory correctly describes
all the momentum levels so long as $g \ll 1$.
%
%


\section{Conclusions}
\label{sec_concl}

We have computed the spectrum of Winter or $\delta$-shell 
model at finite volume both numerically and by means of
perturbative methods.
For very small repulsive couplings,
\beq
0 \, < \, g \, \ll \, \frac{1}{N},
\eeq
the resonance of order $n = 1, 2, 3, \cdots$
of usual Winter model corresponds,
in the finite-volume case, to a quasi-degenerate doublet 
with momenta
\beq
k_n(g) \, = \, n \, - \, g \, n \left( 1 + \frac{1}{N} \right) 
\, + \, \cdots
\qquad 
\mathrm{and} 
\qquad
p_n \, \equiv \, n.  
\eeq
The distance between these levels goes to zero linearly 
with the coupling:
\beq
p_n \, - \, k_n(g) 
\, = \, g \, n \left(1+ \frac{1}{N}\right) \, + \, \cdots.
\eeq
The physical origin of this degeneracy is related
to the fact that, for very small couplings, 
a wavefunction with an integer momentum is,
to a good approximation, an eigenfunction of both the small
$[0,\pi]$ box and the large $[\pi,L]$ one.
In more physical terms, for $g \to 0$,  a particle 
with integer momentum can be entirely
and permanently confined either in the small cavity 
or in the large one.
This degeneracy is well described by ordinary
perturbation theory.

In the large-volume cases,
\beq
N \, \gg \, 1,
\eeq
new approximate degeneracy's occur in the spectrum.
The exact momentum levels $k$
as functions of the coupling $g$, $k=k(g)$, 
show a large variation
of their first derivatives and quasi-degenerate
doublets occur in the spectrum, when
\beq
\label{g_singolari}
g \, \approx \, g_j \, \equiv \, \frac{j}{n \, N};
\qquad j = 1,2,3,\cdots.
\eeq
For generic values of the coupling, of the order
of the inverse of the system size,
\beq
g \, \approx \, \frac{1}{N} \, \ll \, 1,
\eeq 
an infinite-volume resonance corresponds
to the compression of three lines 
in the typical spacing between two of them
and, by increasing further the coupling,
to a mild compression of many lines,
let's say $h=4,5,\cdots$ lines
in the typical spacing of $h-1$ lines 
(see fig.~\ref{fig_pdifferenza}).

The perturbative expansion of the momentum
spectrum of the finite-volume model contains secular
terms, from third order in $g$ on, of the form
\beq
g^i \, N^{i-2}, \quad  g^i \, N^{i-3}, \quad g^i \, N^{i-4}, 
\quad \cdots
\qquad (i = 3,4,5,\cdots).
\eeq
For large volumes, $N \gg 1$, 
these terms can be of order one (or bigger) 
even in the weak coupling regime $g \ll 1$,
so they tend to spoil the convergence of the
perturbative series.
Not surprisingly, ordinary perturbation theory is completely 
unable to describe these approximate degeneracy's
of the spectrum.
By resumming the perturbative series for the
momenta $k = k(g,\xi)$ to all orders in the effective coupling
\beq
\xi \, \equiv \, g \, N \, = \, \mathcal{O}(1),
\eeq
and to some (finite) order in $g \ll 1$,
one obtains improved expansions
presenting pole singularities at $g = g_j$.
We may say that, in some sense, the resummed formulae "see" 
the quasi-degeneracy points $g_j$.
After the exclusion of small neighborhoods
of the singular couplings $g_j$ and proper
matching of the resummed levels with the 
exact ones, 
improved perturbation theory accurately
describes the complete momentum spectrum.
We can therefore claim that
resummed perturbation theory
provides a satisfactory analytic
description of resonances at finite volume 
in the case of the Winter model. 


As far as the physical relevance of 
Winter model at finite volume is concerned,
we believe that this system can provide, 
as discussed in the Introduction, highly 
non-trivial checks of the general superposition 
principle of quantum mechanics.
By means of spectral decomposition,
one finds that the dynamics of a 
standard resonance is 
controlled by the interference of 
its energy eigenstates.
In general, resonance dynamics at finite 
volume involves more complicated superpositions
of energy eigenfunctions than at infinite volume,
because of additional wave-reflection phenomena.
Let us consider an initial wavefunction $\psi_0(x)$ 
with the particle inside the small cavity,
where $\left| \psi_0 \right| = \mathcal{O}(1)$.
At a time
\beq
t \, \approx \, T \, \equiv \, \frac{L}{v},
\eeq
where $L \gg 1$ is the large-cavity size
and $v$ is the average particle velocity,
the wavefunction completely spreads out in the large 
cavity, where it is consequently very small,
\beq
| \psi(T) | \, \approx \, \frac{1}{\sqrt{L}} \, \ll \, 1.
\eeq
Yet, because of coherence, by evolving the system 
by even larger times than $T$, recurrence 
will occur and the particle  
will be back in the small cavity
with a probability close to one.
Eventually, limited-decay phenomena may occur.
If we approximate the $\delta$-function of
Winter model Hamiltonian by means of a 
rectangular potential barrier, the finite-volume model 
can be physically realized by a quantum well%
\footnote{
Because of the existence of the crystal lattice
in the quantum well, the free-particle 
eigenfunctions of the finite-volume Winter model
have to be replaced by Bloch wavefunctions.
}.
In this way, recurrence and limited-decay phenomena
can be experimentally investigated.


\vskip 0.5truecm
\centerline{\bf Acknowledgments}
\vskip 0.5truecm

\noindent
I wish to thank D.$\,$Anselmi and 
M.$\,$Papinutto for discussions.



\end{document}